\documentclass[fleqn,usenatbib]{mnras}

\usepackage{newtxtext,newtxmath}

\usepackage[T1]{fontenc}

\DeclareRobustCommand{\VAN}[3]{#2}
\let\VANthebibliography\thebibliography
\def\thebibliography{\DeclareRobustCommand{\VAN}[3]{##3}\VANthebibliography}

\usepackage{graphicx}	
\usepackage{amsmath}	
\usepackage{pdflscape}

\newcommand{\dotdeg}{\rlap{.}^\circ}

\title[BEBOP-4\,b, eccentric brown dwarf]{BEBOP VIII. SOPHIE radial velocities reveal an eccentric, circumbinary brown dwarf}

\author[A.H.M.J. Triaud et al.]{
Amaury H.M.J. Triaud,$^{1}$\thanks{E-mail: a.triaud@bham.ac.uk}
Thomas A. Baycroft,$^{1}$
Neda Heidari,$^{2}$
Alexandre Santerne,$^{3}$
Aleyna Adamson,$^{1}$
\newauthor
Isabelle Boisse,$^{3}$
Gavin A.L. Coleman,$^{4}$
Alexandre C.M. Correia,$^{5,6}$
Yasmin T. Davis,$^{1}$
Magali Deleuil,$^{3}$
\newauthor
Guillaume Hébrard,$^{2}$
David V. Martin,$^{7}$
Pierre F.L. Maxted,$^{8}$
Richard P. Nelson,$^{4}$
Lalitha Sairam,$^{9}$
\newauthor
Daniel Sebastian,$^{10,1}$
Matthew R. Standing,$^{11}$
Adam T. Stevenson,$^{1}$
Amalie Stokholm,$^{1}$
Mathilde Timmermans,$^{1}$
\newauthor
Stéphane Udry$^{12}$
\\
$^{1}$School of Physics and Astronomy, University of Birmingham, Edgbaston, Birmingham B15 2TT, UK\\
$^{2}$Institut d’astrophysique de Paris, UMR 7095 CNRS universit\'e Pierre et Marie Curie, 98 bis, boulevard Arago, 75014 Paris, France\\
$^{3}$Aix Marseille Univ, CNRS, CNES, LAM, Marseille, France\\
$^{4}$Astronomy Unit, Department of Physics and Astronomy, Queen Mary University of London, Mile End Road, London, E1 4NS, UK\\
$^{5}$CFisUC, Departamento de Fisica, Universidade de Coimbra, 3004-516 Coimbra, Portugal\\
$^{6}$IMCCE, UMR8028 CNRS, Observatoire de Paris, PSL Universit\'e, 77 avenue Denfert-Rochereau, 75014 , Paris, France\\
$^{7}$Department of Physics and Astronomy, Tufts University, 574 Boston Avenue, Medford, MA 02155, USA\\
$^{8}$Astrophysics Group, Keele University, Keele, Staffordshire, ST5 5BG, UK\\
$^{9}$Institute of Astronomy, University of Cambridge, Madingley road, Cambridge, CB3 0HA, UK\\
$^{10}$Th\"uringer Landessternwarte Tautenburg, Sternwarte 5, 07778 Tautenburg, Germany\\
$^{11}$European Space Agency (ESA), European Space Astronomy Centre (ESAC), Camino Bajo del Castillo s/n, E-28692 Villanueva de la Ca\~nada, Madrid, Spain\\
$^{12}$Observatoire Astronomique de l’Université de Genève, Chemin Pegasi 51, CH-1290 Versoix, Switzerland\\
}

\date{Accepted XXX. Received YYY; in original form ZZZ}

\pubyear{\the\year{}}

\begin{document}
\label{firstpage}
\pagerange{\pageref{firstpage}--\pageref{lastpage}}
\maketitle

\begin{abstract}
Circumbinary configurations offer a test of planet formation in an altered environment, where the inner binary has perturbed a protoplanetary disc. Comparisons of the physical and orbital parameters 
between the circumbinary planet population and the population of exoplanets orbiting single stars will reveal how these disc perturbations affect the assembly of planets. Circumbinary exoplanets detected thus far typically have masses $< 3 \,\rm M_{jup}$ raising the question of whether high-mass circumbinary planets are possible, and also whether population features such as the brown dwarf desert would appear in circumbinary configurations like for single star systems. Here, we report observations taken with the SOPHIE high-resolution spectrograph. These observations reveal an $m_{\rm b}\,\sin i_{\rm b} = 20.9 \,\rm M_{jup}$ outer companion, on an eccentric ($e = 0.43$), $1800\,\rm d$ orbit, which we call BEBOP-4\,(AB)\,b. Using dynamical arguments we constrain the true mass $m_{\rm b}< 26.3 \,\rm M_{jup}$. The inner binary's two eclipsing stellar components have masses $M_{\rm A} = 1.51\,\rm M_\odot$, and $M_{\rm B} = 0.46\,\rm M_\odot$. Their orbital period is $72\,\rm d$, and their eccentricity is $0.27$. This system contains the longest period binary surveyed by the BEBOP project. 
BEBOP-4\,b is expected to be detectable using {\it Gaia} DR4 single epoch astrometric measurements. Despite a large period ratio of $\sim 25:1$, the substantial eccentricities of both orbits mean that the outer orbit is on the edge of orbital stability, and located in between two destabilising secular resonances. Should the outer companion survive, the BEBOP-4 system appears like a precursor to several post-common envelope binaries exhibiting eclipse timing variations where very massive circumbinary companions have been proposed.
\end{abstract}

\begin{keywords}
exoplanets -- binaries: eclipsing -- brown dwarfs -- radial velocities
\end{keywords}



\section{Introduction}\label{sec:intro}
Only few circumbinary exoplanets are known. 14 have been discovered thanks to transits \citep[e.g.][]{doyle_kepler-16_2011, orosz_kepler-47_2012, kostov_toi-1338_2020}, whilst another three have been identified from radial-velocity observations \citep{standing_radial-velocity_2023,baycroft25a, baycroft25b}. In addition, nine have been proposed using direct imaging \citep[e.g.][]{Bailey2014,Dupuy2018}, five from microlensing \citep[e.g.][]{Bennett2016,Han_2024}, one has been confirmed from eclipse timing variations \citep{goldberg_5mjup_2023}, with another 38 being proposed \citep[e.g.][]{Borkovits2015, Beuermann2010, baycroft_new_2023,Basturk_2023}, and one has been detected from pulsar timing variations \citep{Thorsett_1993,Sigurdsson_2003}. Compared to over 6,000 exoplanets known to orbit single stars, circumbinary planet would appear rare at first sight. However, about $10-12\%$ of close binaries host circumbinary gas giants \citep{martin_planets_2014, Armstrong14,Baycroft24}, a number which is similar to single stars \citep{Cumming2008, Mayor2011, Santerne2016, Fulton2021}.

Another rare configuration is a single Sun-like star with a brown dwarf companion \citep[e.g.][]{Grether2006, Sahlmann2014, Ma2014, triaud_eblm_2017, Carmichael2019,Dalal2021}. Dubbed the {\it brown dwarf desert}, observations reveal that, particularly on orbital periods $< 100-1000~\rm days$ \citep{Kiefer2019}, very few brown dwarfs exists, with the rarest being of a mass in the range $25-45\,\rm M_{jup}$ \citep{Sahlmann2011,Stevenson2023}. This pattern has been interpreted for companions nearer the Hydrogen-burning limit \citep[$\sim 80\,\rm M_{jup}$;][]{Kumar1963,Baraffe2015} as the tail-end of stellar formation, while near the Deuterium-burning limit \citep[$\sim 13\,\rm M_{jup}$;][]{Baraffe2003} those objects might represent the high-end of planet formation, whether from core-accretion followed by runaway gas accretion \citep{Pollack1996,coleman_global_2023}, or via direct collapse \citep{StamatellosWhitworth2009}. As a result, a $25\rm \,M_{jup}$  has been adopted to separate between planetary-mass and stellar-mass brown dwarfs \citep{Schneider2011, Sahlmann2011}.

Whether circumbinary brown dwarfs exist remains a puzzle. In two systems detected with direct imaging, Delorme-1 \citep{Delorme13} and VHS\,J1256-1257 \citep{Dupuy23}, the companion mass is of the same order of magnitude as the components of the binary \citep[possibly also in 2M1510;][]{Calissendorff19, baycroft25a}, hence giving the impression they are closer to a triple system than a true circumbinary configuration. Furthermore, in those systems, as a well as in all other directly detected circumbinary brown dwarfs, the brown dwarf is found at a large separation from the inner binary, far from the short-period brown dwarf desert seen in single Sun-like stars. The most evidenced short period circumbinary brown dwarf is in HD\,202206, however its nature is debated. This system contains a Sun-like star with two companions, one with $m_{\rm b} \sin i_{\rm b} = 17\,\rm M_{jup}$, and $m_{\rm c} \sin i_{\rm c} = 2.4\,\rm M_{jup}$ \citep{correia_coralie_2005}. Whilst \citet{Couetdic2010} use orbital stability arguments to estimate the true masses are $m_{\rm b} <34\,\rm M_{jup}$ and $m_{\rm c} <4.4\,\rm M_{jup}$, and thus $i_{\rm b} \sim i_{\rm c} > 30^\circ$, \citet{benedict_hd_2017} argue instead that astrometric measurements made with {\it Hubble}'s fine guidance sensor imply $i_{\rm b} \sim i_{\rm c}\sim 8-10^\circ$, which would mean $m_{\rm b} = 0.09\,\rm M_\odot$ and $m_{\rm c} = 18\,\rm M_{jup}$, a circumbinary brown dwarf. Eventually {\it Gaia} DR4 astrometry should resolve this conflict, but the orbital stability arguments are hard to shake and more likely than not, \citet{Couetdic2010} are correct and HD\,202206 is likely not hosting a circumbinary brown dwarf.

Meanwhile, BEBOP (Binaries Escorted By Orbiting Planets) is a project to survey nearby close binaries with radial-velocities to detect circumbinary exoplanets. The project started in 2014, with a large sample of low-mass eclipsing binaries, and a limited radial-velocity precision. In the first publication of the project, we could place an upper limit on the occurrence rate of circumbinary brown dwarfs $\eta < 6.5\%$ \citep{martin_bebop_2019}. In more recent work, we also notice a paucity of companions with mass $> 3 \,\rm M_{jup}$, found to be $5\times$ rarer than in equivalent single star systems \citep{Baycroft24}. In fact, within the confirmed transit, radial-velocity and eclipse-timing population, only one circumbinary planet has a mass $> 3 \,\rm M_{jup}$, Kepler-1660\,b, with $4.9  \,\rm M_{jup}$ \citep{goldberg_5mjup_2023}.

In this paper, we now report the clear detection of a circumbinary brown dwarf in a system we call BEBOP-4. This discovery is important to better understand planet formation. Already planetary-mass brown dwarfs are difficult to assemble in single star systems and they are very rarely obtained in simulations of planetary system formation \citep[e.g.][]{Coleman2016, Emsenhuber2021}. However, circumbinary planet formation is expected to be more difficult yet, at least in the close vicinity of the central binary, since the circumbinary protoplanetary disc is anticipated to be highly disturbed and more turbulent than for a single star. Vigorous turbulence in protoplanetary discs typically reduces the efficiency of pebble accretion \citep[e.g.][]{Rosenthal2018,pierens_vertical_2021}, and perturbations to planetesimal orbits also slow the rate of planetesimal accretion \citep{paardekooper_how_2012}.

In section \ref{sec:data} we describe our data collection, and how radial-velocities are extracted from the spectra. In section \ref{sec:M1}, we discuss how we measure the masses for both components of the stellar binary. These data are modeled in section \ref{sec:res}. We interpret and discuss our results in section \ref{sec:dis} before concluding.

\begin{figure*}
    \centering \includegraphics[trim={0.0cm 0 0.0cm 0},clip=true,width=0.295\linewidth]{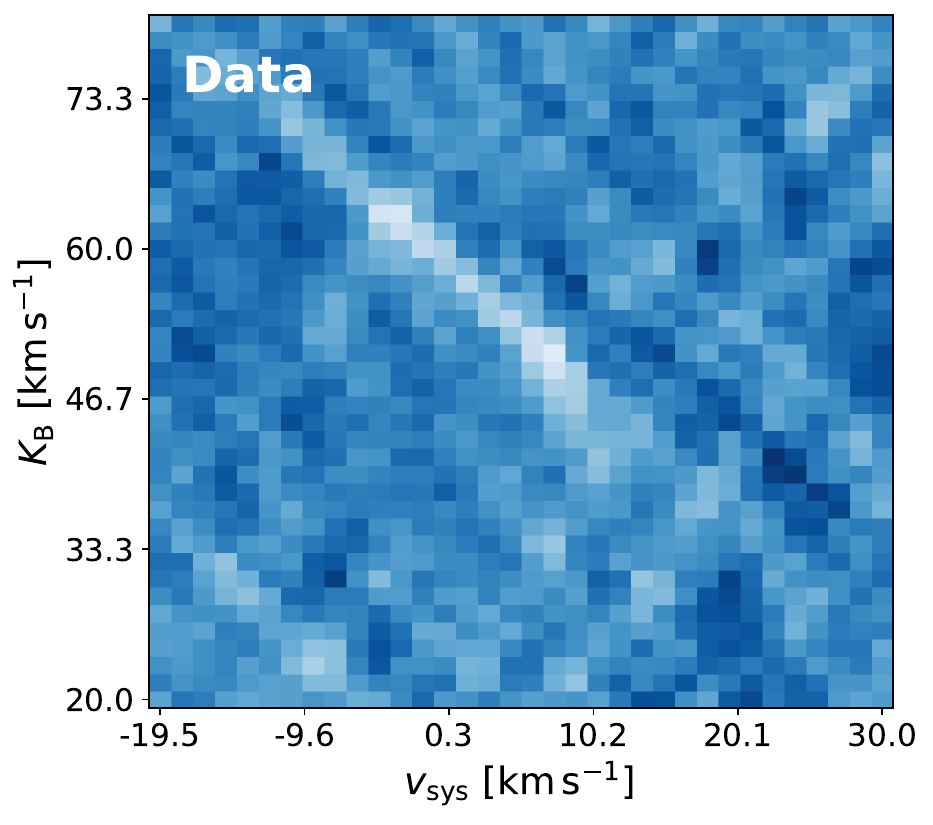}\hspace{-0.1cm}\includegraphics[trim={0.0cm 0 0.0cm 0},clip=true,width=0.295\linewidth]{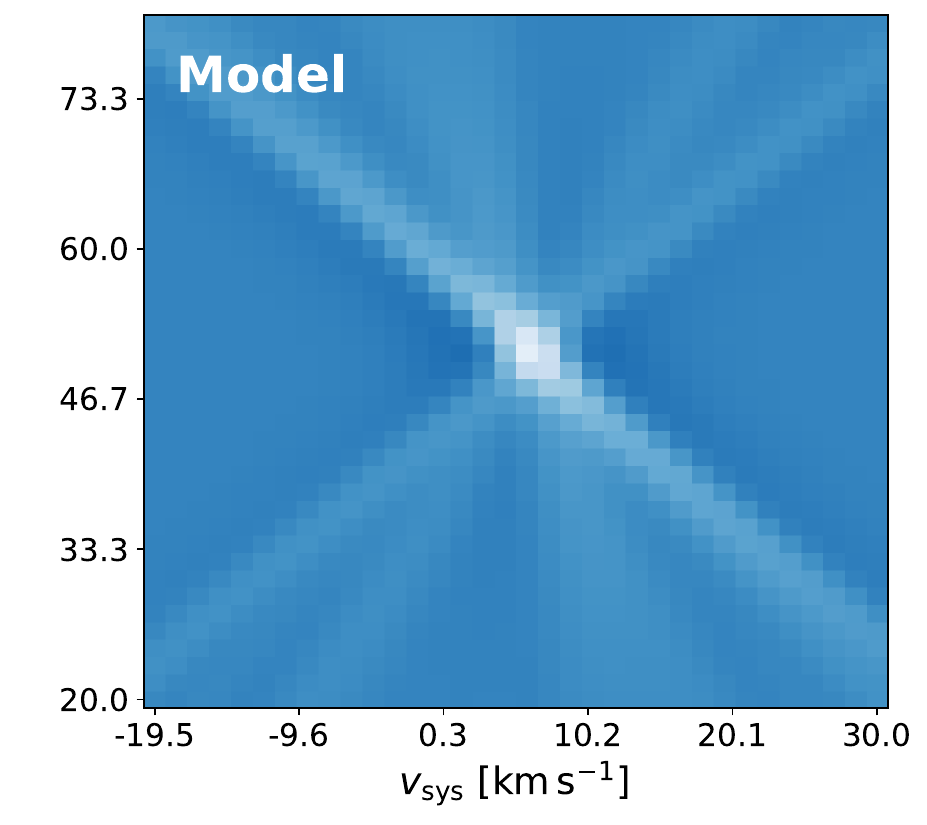}\hspace{-0.1cm}\includegraphics[trim={0.0cm 0 0.0cm 0},clip=true,width=0.39\linewidth]{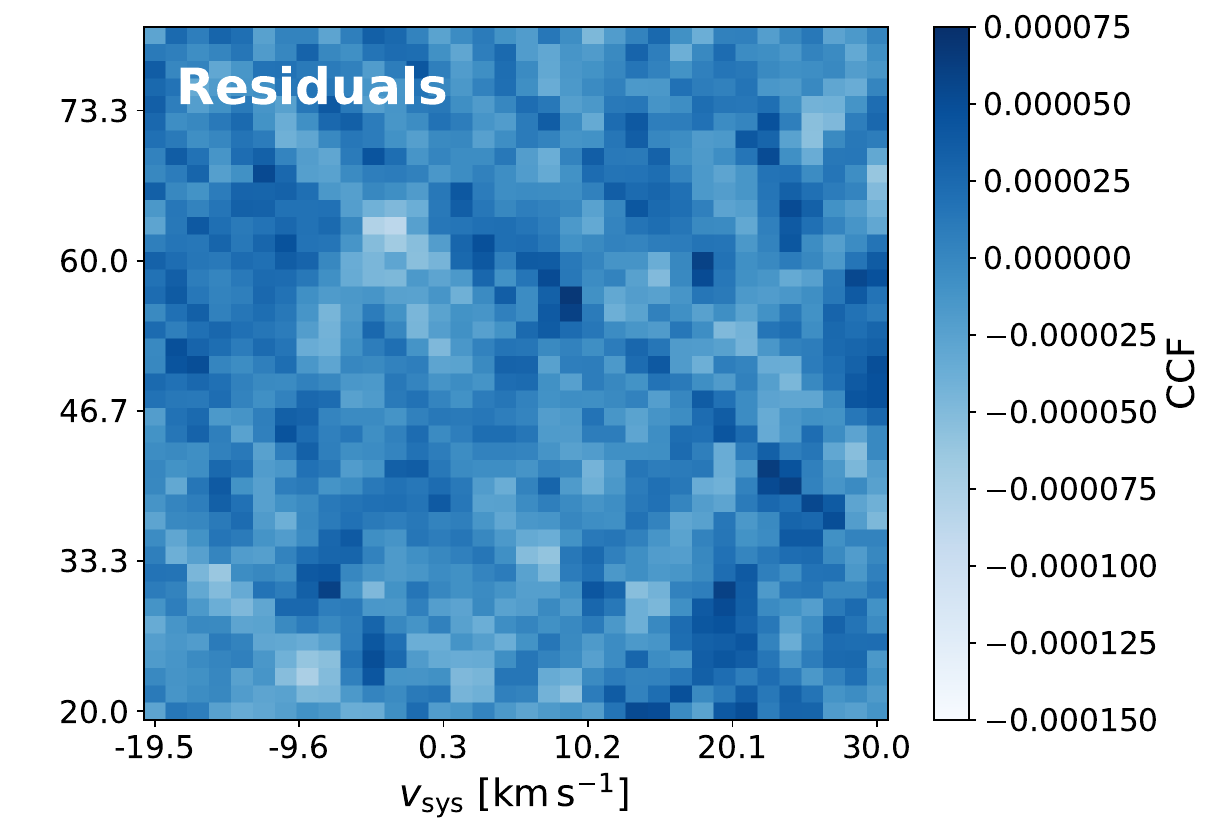}
    \caption{HRCCS detection of BEBOP-4\,B. Left: $K$-focused cross-correlation map. Center: \texttt{Saltire} model for the SOPHIE data, with parameters from MCMC. Right: residual CCF map after subtraction of the model. Data and model have been median-subtracted to allow for a common scale to be shared between the three panels. A secondary maximum is identified in the CCF map with a similar significance at $K_{\rm B}\sim62$\,km\,s$^{-1}$, $v_{\rm sys}\sim-3$\,km\,s$^{-1}$. This false signal, at the incorrect systemic velocity, is actually less significant and only gets highlighted due to the presence of the main signal. In the residual map, the `false' maximum becomes almost indistinguishable from other noise structures.}
    \label{fig:saltire}
\end{figure*}

\section{Radial velocity data collection}\label{sec:data}

Target selection for the Southern component of BEBOP is described in \citet{martin_bebop_2019}. For the Northern survey, a brief description is provided in \citet{baycroft25b}. The system, HD\,111605 ($\alpha$= 12:50:18.53, $\delta$=	+26:09:34.29, V$_{\rm mag} =9.6$), with spectral type G3, was identified by the KELT survey \citep[Kilodegree Extremely Little Telescope;][]{pepper_kilodegree_2007}, and reported as a likely eclipsing binary \citep[][where it is known as KC08C10921]{collins_kelt_2018}, a false positive transiting planet signal. Its properties made it compatible with an EBLM \citep[Eclipsing Binary -- Low Mass;][]{triaud_eblm_2013,maxted_eblm_2023}, like all systems observed during the first phase of the BEBOP radial-velocity search for circumbinary exoplanets \citep{martin_bebop_2019}. As the BEBOP-North survey catalogue was assembled, this system received the designation EBLM~J1250+26 under which name it was observed. The Northern survey started with a short reconnaissance phase (2018-09-01 to 2019-02-28) using the SOPHIE spectrograph \citep{perruchot_sophie_2008} at the Observatoire de Haute Provence (OHP), to assess the radial-velocity precision of each primary star, verify the orbital and physical parameters of the secondaries were in line with expectations, and to remove systems with high stellar-activity \citep[e.g. from a lack of variations in the bisector span as in][]{Queloz2001}. EBLM~J1250+26 was selected as a top priority system for the BEBOP-North survey. Intense radial-velocity monitoring with SOPHIE, for BEBOP-North, started in earnest in 2019-03-11 and are still ongoing.

By construction, the BEBOP survey runs a blind survey, meaning all systems are in principle observed at the same cadence and under similar conditions. The collected data is analysed every six months, at the end of each observing semester to perform a quality control (remove systems with high stellar activity, triple systems etc.), and search for candidates. When candidates are identified, they are flagged, but not observed particularly more intensively so as to avoid introducing an observer's bias, however reporting candidates remains important as a means to convince telescope time allocation committees about the validity of continuing the survey for competitively obtained telescope time.

In the case of EBLM~J1250+26, around March 2020 we had noticed a downward, quadratic drift in the residuals of the orbital motion of the binary. At that time we suspected a triple system, but still within planetary range. The COVID-19 pandemic started and observatories all over the world closed. The OHP stopped operations entirely between mid-March and mid-May 2020, reopening quite early, but with limited availability.
As a consequence, priority was reduced on EBLM~J1250+26 since drifts can be followed efficiently at a slower cadence. Four measurements were obtained, in 2020-05-29, 2020-12-02, 2021-02-25 and 2021-03-03, the period during which travel to OHP was difficult. When data were analysed, it became immediately apparent the last three of these measurements did not follow the expected drift, and that a companion was likely undergoing a periastron passage. An attempt to observe was made during twilight in early August 2020, but by then the target was too low on the horizon. Observations resumed at high cadence and with high priority in 2021-11-25. Regular fits to the orbit indicated a full orbital phase would be completed around 2024, which indeed happened. A new periastron passage is expected in September 2025 when the star will unfortunately be below the horizon at night.

Because of these unique circumstance, compared to other systems of the BEBOP survey, EBLM~J1250+26 (BEBOP-4) is the system that has been the least blindly monitored.

As of 2025-03-01, we collected 84 high resolution spectra, obtained between 2019-01-07 and 2025-02-27, with the high resolution \'echelle spectrograph SOPHIE, mounted on the T193 at the Observatoire de Haute-Provence \citep{perruchot_sophie_2008}. Data were acquired under Prog.ID 18B.DISC.TRIA (PI Triaud) and 19A.PNP.SANT (PI Santerne). All data have been acquired in ObjAB mode where fibre A is placed on the target and fibre B is positioned on the sky to enable the removal of solar spectral lines reflected by the Moon. The observations were acquired using the high resolution (HR) mode with resolution \(R=75\,000\). Standard calibrations were taken at the start of the night and at approximately 2~hr intervals throughout the night to track the instrumental zero-point.

The SOPHIE radial velocities (RVs) are computed by the SOPHIE Data Reduction System \citep[DRS,][]{bouchy_sophie_2009}, which cross-correlates the spectra with a weighted numerical mask corresponding to a G2-type star and subsequently fits a Gaussian to the cross-correlation function (CCF) \citep{pepe_coralie_2002,baranne_elodie_1996}. To enhance the precision of the SOPHIE RV measurements, we applied the recently optimised procedures described in \cite{heidari_overcoming_2022} and \cite{heidari_sophie_2024}. These include: (1) corrections for charge transfer inefficiency in the CCD \citep{bouchy_charge_2009}; (2) correction of template (colour) effects \citep{heidari_sophie_2024}; (3) removal of moonlight contamination by utilising the simultaneous sky spectrum recorded via the second SOPHIE fiber \citep{pollacco_wasp-3b_2008}; (4) correction for long-term instrumental drifts using the RV master constant time series, constructed from so-called ‘constant’ stars monitored each night by SOPHIE \citep{courcol_sophie_2015,heidari_overcoming_2022,heidari_sophie_2024}; and (5) correction for short-term instrumental drifts by interpolating drift measurements at the exact time of each observation, based on frequently measured drifts throughout the night. Radial velocity data are collated into Table \ref{tab:rv_data}.

The 84 spectra obtained have a median SNR = 46.8 (at 5550 \r{A}) and a median radial velocity precision of 3.4 \({\rm m\,s^{-1}}\).

\section{Determining stellar masses}\label{sec:M1}

We could assume a mass of $M_{\rm A} = 1.584\pm0.086\,\rm M_\odot$ for the primary star, as was obtained from a homogeneous spectral analysis of the BEBOP sample presented in \citet{freckelton_bebop_2024}. However, recently developed methods allow some of our binaries to be transformed from a single to a double-lined binary \citep{sebastian_eblm_2024,sebastian2025-kepler16}, to measure absolute, dynamical masses for both components.

We use the same high resolution cross-correlation spectroscopy (HRCCS) method as in \citet{sebastian_eblm_2024}, \citet{sebastian2025-kepler16}, and \citet{baycroft25b}, using Principal Component Analysis (PCA) and cross-correlation in the secondary's rest frame (with the HARPS M2 mask) to extract the weak lines of the secondary star. For this system, we make an additional modification whereby any additional RV shifts caused by the circumbinary brown dwarf companion (Section~\ref{sec:res}) are first removed from the spectra before we attempt to detect the secondary star. 

The CCF maps (Fig.~\ref{fig:saltire}) are fit with the \texttt{Saltire} model \citep{sebastian_saltire_2024} to extract $K_{\rm B}$ and $v_{\rm sys}$. However, the \texttt{Saltire} uncertainties are on the fit parameters, and do not estimate systematic uncertainty from the presence of correlated noise. To quantify the systematic error, we use a bootstrapping analysis, where we randomly draw 50~per~cent of the data and measure the best fitting parameters \citep[as in][]{baycroft25b}. This is repeated 200 times, and we impose that no more than 15 (out of 42) spectra can be common between any permutation \citep[adapted slightly from the recommendations of][to allow 200 random permutations to be generated]{sebastian2025-kepler16}. From the 200 fit results, we take the $1$-$\sigma$ percentiles around the median as the systematic uncertainty in the parameters. The MCMC fit error is $\pm0.207\,\rm{km\,s}^{-1}$, and the systematic error is $\pm2.02\,\rm{km\,s}^{-1}$, and these are added quadrature.

We detect a signal with 5.8-$\sigma$ significance, and measure a secondary semi-amplitude of $K_{\rm B} = 51.02 \pm 2.03\,\rm km\,s^{-1}$, implying $M_{\rm A} = 1.507\pm 0.152$ and $M_{\rm B} = 0.459\pm 0.028$ (assuming $i_{\rm bin}=\mathcal{U}[88.5,91.5]^{\circ}$, and following the IAU recommended equations; \citealt{prsa2016}). 

Unfortunately no primary eclipses were observed by {\it TESS} \citep{ricker_transiting_2015} so we cannot perform an analysis like in \citet{davis_eblm_2024}, to measure the stellar radii accurately. This analysis will be performed once a primary eclipse has been collected by {\it TESS}, although no new observations of this target are scheduled through {\it TESS} Cycle 8.

We note that $M_{\rm A}$ is measured differently by different methods (see Fig.~\ref{fig:mass-comparison}). Thanks to our new measurement, we confirm the fiducial $M_{\rm A}$ masses used by the BEBOP survey as produced in \citet{freckelton_bebop_2024} are fully consistent to within 1-$\sigma$ for this system. Using HRCCS, we likely measure $M_{\rm A}$ more accurately, though at a slightly reduced precision to that achieved through spectral analysis and isochrone fitting, owing to the relatively low detection significance of the secondary.

\section{Analysis and results}\label{sec:res}


The radial-velocities were analysed with {\tt kima} \citep{faria_kima_2018}, using the "BINARIESmodel" \citep{baycroft_improving_2023}. The usual orbital elements are used as parameters. In addition, we fit a {\it jitter} term to rescale the uncertainties should they need to. The model also has an option to fit for apsidal precession, by having one additional free parameter, $\dot \omega$, and also includes a radial velocity correction due the effects from general relativity \citep{baycroft_improving_2023, baycroft25a}. We chose to let $\dot \omega$ be a free parameter. A non-zero value would provide an additional constraint on the companion's parameters. This model, and the fitting procedure have been validated and used in several publications \citep{baycroft_improving_2023, standing_radial-velocity_2023, sairam_new_2024, Sairam2024b, baycroft25a, baycroft25b}.

\begin{figure}
    \centering
    \includegraphics[width=1\linewidth]{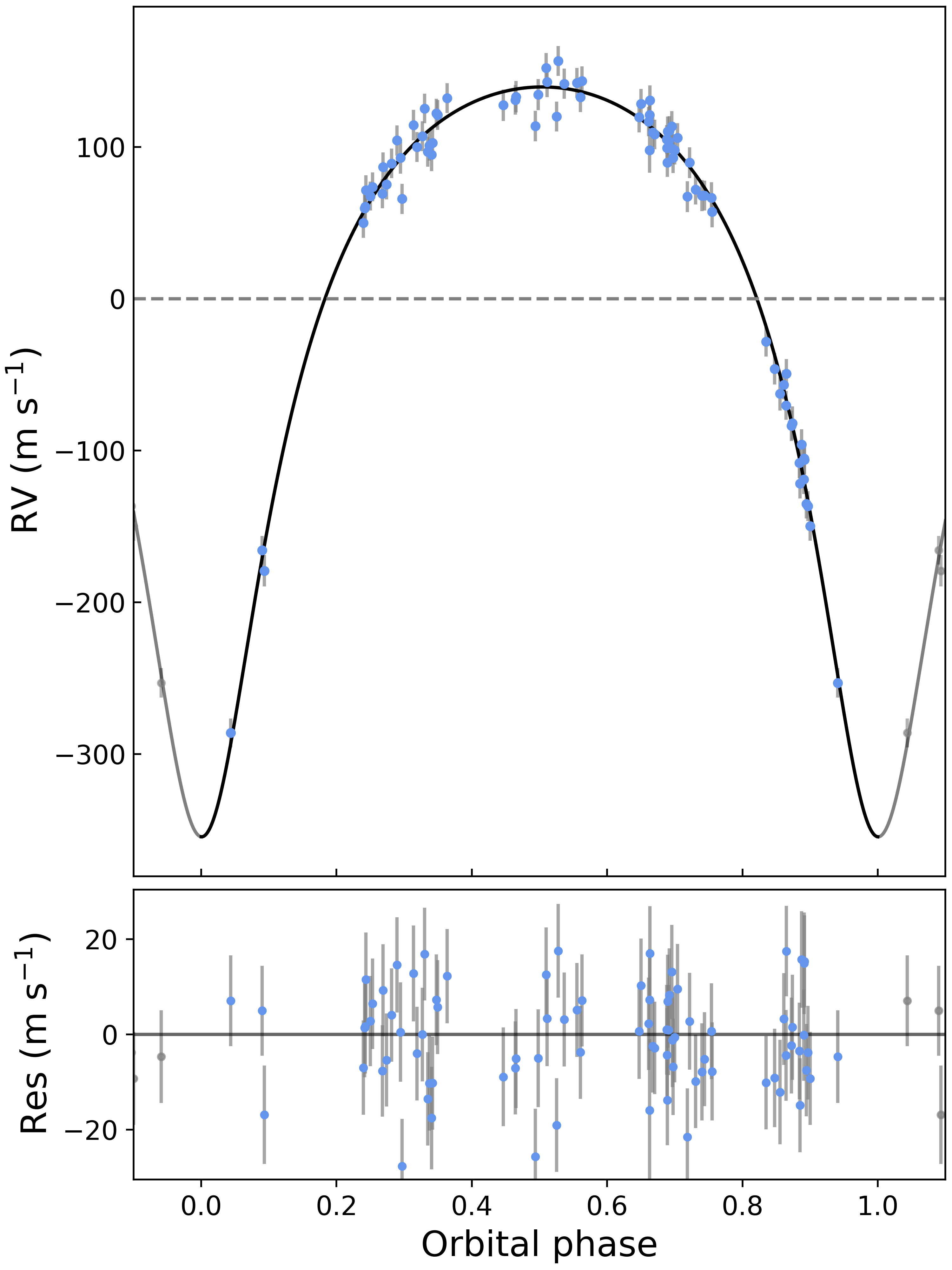}
    \caption{Phased radial velocity Keplerian and residuals for the detected circumbinary companion. The parameters used are those from the maximum-likelihood solution.}
    \label{fig:rvcurve}
\end{figure}

\begin{figure}
    \centering
    \includegraphics[trim=5cm 1cm 2.5cm 2cm,width=\linewidth]{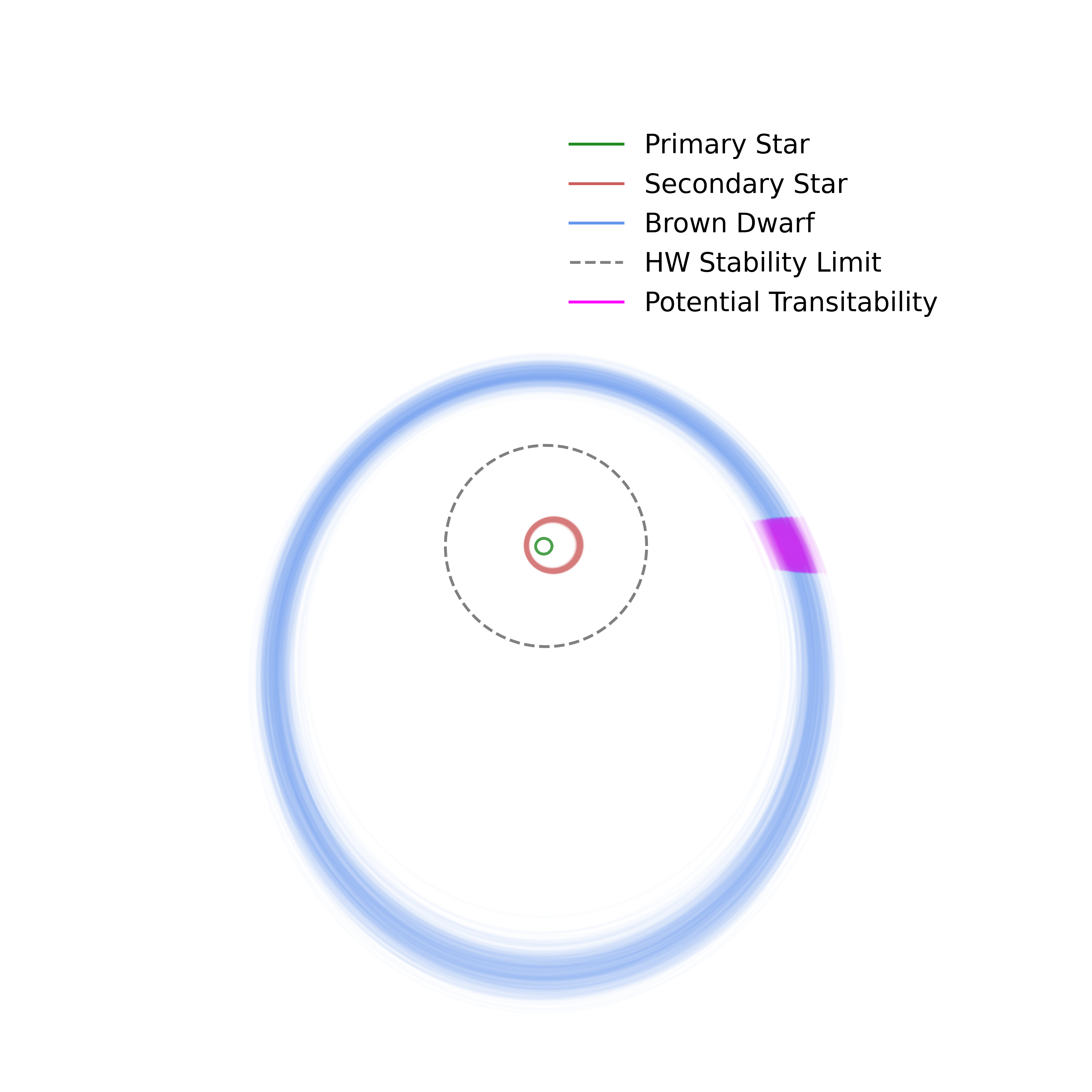}
    \caption{ Face-on view of the orbital configuration of the BEBOP-4 system showing the orbits of each star and the brown dwarf companion for 600 posterior draws. The observer is located on the right hand side. For illustrative purposes, the stability limit is calculated using \citep{holman_long-term_1999} and the binary orbital parameters from the maximum-likelihood solution is also shown. In magenta we highlight the parts of the companion orbit where it might transit one of the stars  (calculated as the orbital phases where the companion crosses in front of the area swept out by the binary orbit, assuming coplanarity).}
    \label{fig:orbit}
\end{figure}

In our resulting fit, we find a model with a circumbinary companion is clearly preferred with a Bayes Factor $BF \sim 10^{64}$ over a model with no circumbinary companion. The best fitting result for the companion and the binary, results in residuals with RMS scatter of $ 10.3 \,\rm m\,s^{-1}$, this is larger than the mean uncertainty of the dataset and requires a jitter of $\approx 10$ $\rm m\,s^{-1}$. The companion has an orbital period $P_{\rm b} = 1823.5^{+5.1}_{-4.9}\,\rm d$, and a semi-amplitude $K_{\rm b} =243.7\pm6.5 \rm m\,s^{-1}$. All fit parameters are reported in Table~\ref{tab:parameters}. The radial velocity curve for the companion is shown in Fig. \ref{fig:rvcurve}.  The overall highest likelihood model is plotted in Fig.~\ref{fig:binary_wdot} and the binary Keplerian in Fig.~\ref{fig:binary_rv}. The posterior on the orbital parameters of the outer companion are graphically represented in Fig.~\ref{fig:planet_corner}. The orbital configuration of the system is shown in Fig. \ref{fig:orbit}, where 600 random draws from the posterior are plotted. The region shown in magenta is where the companion might pass in front of the binary and as such a transit could potentially happen if the configuration is right.

We use $M_{\rm A}$ and $M_{\rm B}$ from Section~\ref{sec:M1}, and thus the minimum mass of the circumbinary companion is
$m_{\rm b}\sin{i_{\rm b}} = 20.9\pm1.3\rm \,M_{jup}$, above the Deuterium-burning limit \citep{Kumar1963,Baraffe2003}, but below the $25\rm \,M_{jup}$ that have been taken as the separation between planetary-mass and stellar-mass brown dwarfs \citep{Schneider2011, Sahlmann2011}.  Constraints on the mutual inclination between the outer brown dwarf's orbit and the inner eclipsing binary can be obtained with a measurement of the binary's apsidal precession.

\begin{table}
    {\centering
    \caption{Orbital parameters for the BEBOP-4 binary, planet, and general fit parameters. Stellar masses, semi-major axes, and $K_{\rm B}$ are calculated from SB2 analysis in Section~\ref{sec:M1} as well as the {\tt kima} analysis, and all remaining parameters are taken or derived from the {\tt kima} analysis of RVs. Note that these are mean orbital elements rather than osculating elements. Since $\omega$ evolves with time, the reference time at which the value reported is calculated is 2\,459\,736.088894 (BJD). This reference time is also that at which \(\mathcal{M}_0\) and \(\lambda_0\) are calculated. The spectral type and a variety of alternate names for the BEBOP-4 system are listed as well. }
    \begin{tabular}{l|c|c}
        \hline
        \multicolumn{3}{c}{BEBOP-4 -- G3\,V}\\
        \multicolumn{3}{c}{$\alpha = 12^{\rm h}50^{\rm m}18.52^{\rm s}; \delta = +26^\circ09'34.0"$}\\
        \hline
        \multicolumn{3}{c}{HD 111605; Gaia DR3 3961485079994112896;}\\
        \multicolumn{3}{c}{EBLM\,J1250+26; TIC 356710041; TOI-2065;}\\
        \multicolumn{3}{c}{UCAC4 581-047798; 2MASS\,J12501852+2609340}\\        
        \hline
        Parameter & Units & Value \\
        \hline
        \multicolumn{3}{c}{\textit{Orbital fit parameters}}\\
        & & \\
        \(P_{\rm bin}\) & [days] & \(71.96857\pm0.00015\)\\
        \(K_{\rm A}\) & [\({\rm m\,s^{-1}}\)] & \(15\,547.3\pm2.1\)\\
        \(K_{\rm B}\) & [\({\rm m\,s^{-1}}\)] & \(51\,018\pm2031\)\\
        \(e_{\rm bin}\) & & \(0.26887\pm0.00011\) \\
        \(\omega_{\rm bin}\,^{\rm (a)}\) & [rad] & \(4.81629^{+0.00049}_{-0.00051}\) \\
        \(\dot{\omega}_{\rm bin}\) & [\({\rm ''\,yr^{-1}}\)] & \(88^{+49}_{-46}\) \\
        \(\mathcal{M}_{0,\rm bin}\) & [rad] & \(1.00142^{+0.00048}_{-0.00043}\) \\
        & & \\
        \(P_{\rm b}\) & [days] & \(1823.5^{+5.1}_{-4.9}\) \\
        \(K_{\rm b}\) & [\({\rm m\,s^{-1}}\)] & \(243.7\pm6.5\) \\
        \(e_{\rm b}\) & & \(0.428\pm0.016\) \\
        \(\omega_{\rm b}\,^{\rm (a)}\) & [rad] & \(3.134\pm0.020\) \\
        \(\mathcal{M}_{0,\rm b}\) & [rad] & \(2.162\pm0.021\) \\
        \hline
        \multicolumn{3}{c}{\textit{Assumed and derived parameters}}\\
        & & \\
        \(i_{\rm bin}\) & [rad] & \([1.545,\pi/2]\) \\
        \(M_{\rm A}\) & [\({\rm M_{\odot}}\)] & \(1.507\pm0.152\) \\
        \(M_{\rm B}\) & [\({\rm M_{\odot}}\)] & \(0.459\pm0.028\) \\
        \(a_{\rm A}\) & [AU] & \(0.099063\pm0.000013\)\\
        \(a_{\rm B}\) & [AU] & \(0.325\pm0.013\)\\
        \(\lambda_{0, \rm bin}\,^{\rm (a)}\) & [rad] & \(6.34386\pm0.00030\) \\
        \({T_{\rm peri,bin}}\) & [BJD-2\,400\,000] & \(59\,724.6183^{+0.0049}_{-0.0056}\) \\
        \(i_{\rm b}\) & [rad] & \([0.917,\pi/2]\) \\
        \(m_{\rm b}\sin{i}_{\rm b}\) & [\({\rm M_{jup}}\)] & \(20.9\pm1.3\) \\
        \(a_{\rm b}\) & [AU] & \(3.634\pm0.011\) \\
        \(\lambda_{0, \rm b}\,^{\rm (a)}\) & [rad] & \(5.820\pm0.017\) \\
        \({T_{\rm peri,b}}\) & [BJD-2\,400\,000] & \(59\,108.5\pm5.7\) \\
        \hline
        \multicolumn{3}{c}{\textit{Other fit parameters}}\\
        & & \\
        Jitter & [\({\rm m\,s^{-1}}\)] & \(10.0\pm1.1\) \\
        \(\nu\) & (Student's t shape) & \(74^{+354}_{-62}\) \\
        \(v_{\rm sys}\) & [\({\rm m\,s^{-1}}\)] & \(5277.5\pm1.8\)\\
        \hline
    \end{tabular}\\}
    \(^{\rm (a)}\) The values of the argument of pericentre \(\omega\) and the true longitude at the refernce time \(\lambda_0\) reported are those of the orbit of the primary star around the centre-of-mass of its respective two-body orbit, this is the \(\omega\) used in the radial velocity equation. To obtain the \(\omega\) or \(\lambda_0\) of the secondary star's or brown dwarf's orbit, \(\pi\) should be subtracted from the values reported.
    \label{tab:parameters}
\end{table}

The apsidal precession of the BEBOP-4(AB) binary is constrained to $\dot{\omega}_{\rm bin} =88^{+49}_{-46} {\rm\,''\,yr^{-1}}$. The contribution from general relativity\footnote{The apsidal precession induced by tidal deformation and rotational flattening is negligible.} \citep[using equation A17 from][]{baycroft_improving_2023} is $\dot{\omega}_{\rm bin} = 0.977\pm0.059 {\rm\,''\,yr^{-1}}$, meaning that the measured apsidal rate that would be induced by a tertiary companion is $\dot{\omega}_{\rm bin} = 87^{+49}_{-46} {\rm\,''\,yr^{-1}}$. Figure \ref{fig:binary_wdot} depicts the contribution of $\dot{\omega}_{\rm bin}$ to the fit. We use \(\Delta i\) to represent the difference between the binary's and brown dwarf's inclinations, analogous to the mutual inclination under the assumption that the longitudes of ascending nodes of the two orbits are equal. Assuming the brown dwarf companion is coplanar with the binary (i.e. \(\Delta i = 0^{\circ} \Rightarrow m_{\rm b} = m_{\rm b}\sin{i_{\rm b}} = 20.9\pm1.3 \rm \,M_{jup}\)), the predicted apsidal precession \citep[using equation A5 from][]{baycroft_improving_2023} is $\dot{\omega}_{\rm bin} = 101.2\pm5.4 {\rm\,''\,yr^{-1}}$, consistent with the measured value. 

\begin{figure}
    \centering
    \includegraphics[width=\linewidth]{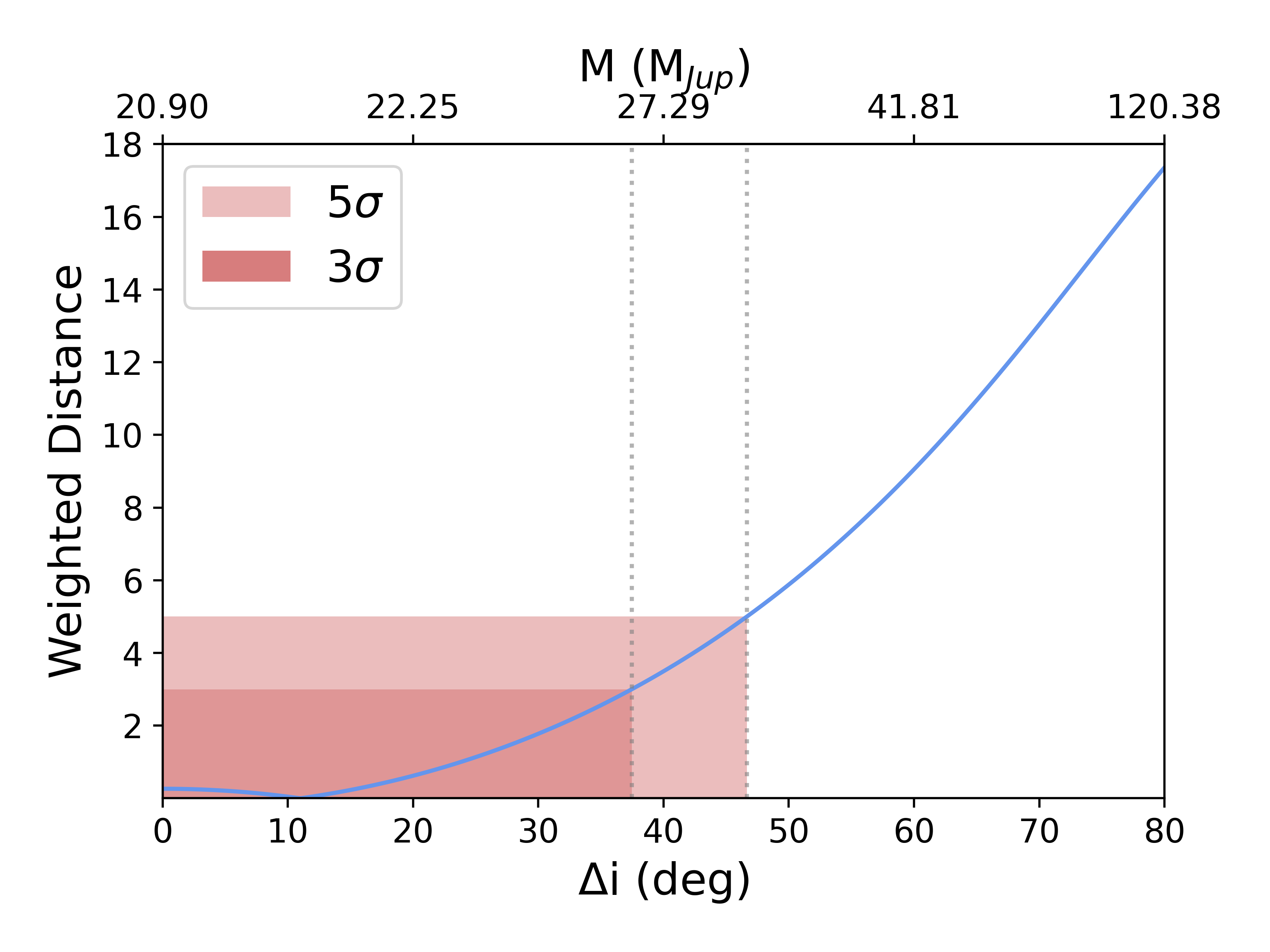}
    \caption{Weighted distance between the measured apsidal precession rate (with the GR contribution removed) and the predicted apsidal precession for various mutual inclination values (i.e. inclinations of the circumbinary companion). Dashed lines show the $3\sigma$ and $5\sigma$ constraints.}
    \label{fig:wdotinc}
\end{figure}

We can use the apsidal precession constraint to obtain an upper limit on \(\Delta i\) and therefore also on the true mass of the companion. As the \(\Delta i\) increases, the induced apsidal precession rate decreases and eventually becomes negative as the companion moves from a coplanar to a polar orbit \citep{Zhang_2019,baycroft25a}. The weighted distance between the measured precession rate and the predicted precession rates as a function of inclination is shown in Fig. \ref{fig:wdotinc}, along with the corresponding masses. At\footnote{Here we assume the inclination of the binary \(i_{\rm bin} = \pi/2\).} $3\sigma$, the mutual inclination is \(\Delta i < 37.5^{\circ}\), \( 0.917 \leq i_{\rm b} \leq2.225 \rm \,(rad)\), and $m_{\rm b} \leq 26.3\,\rm M_{jup}$. At $5\sigma$: \(\Delta i < 46.7^{\circ}\), \( 0.756\leq i_{\rm b} \leq2.385 \rm \,(rad)\), and $m_{\rm b} \leq 30.5\, \rm M_{ jup}$\footnote{The function converting $\dot{\omega}$ to $\Delta i$ is symmetrical about $\Delta i=90^\circ$. As such retrograde orbits are also possible.}. This constrains BEBOP-4b to being in the planetary-mass regime of brown dwarfs, as such it enters the BEBOP catalogue of circumbinary planets shown in Table \ref{tab:catalogue}. This constraint also indicates that the outer orbit's mutual inclination is most likely below the classical Lidov-Kozai angle \citep[$39\dotdeg21$;][]{Lidov1962,Kozai1962}. We also note that the BEBOP-4 configuration is tight enough that Lidov-Kozai cycles induced by the companion on the inner binary would be unlikely anyway. Lidov-Kozai cycles happen when the ratio of angular momenta $L_{\rm out} / L_{\rm in} > 0.5$ \citep{Martin2016} and here $L_{\rm b} / L_{\rm bin} =0.16$. 

\begin{figure}
    \centering
    \includegraphics[width=\linewidth]{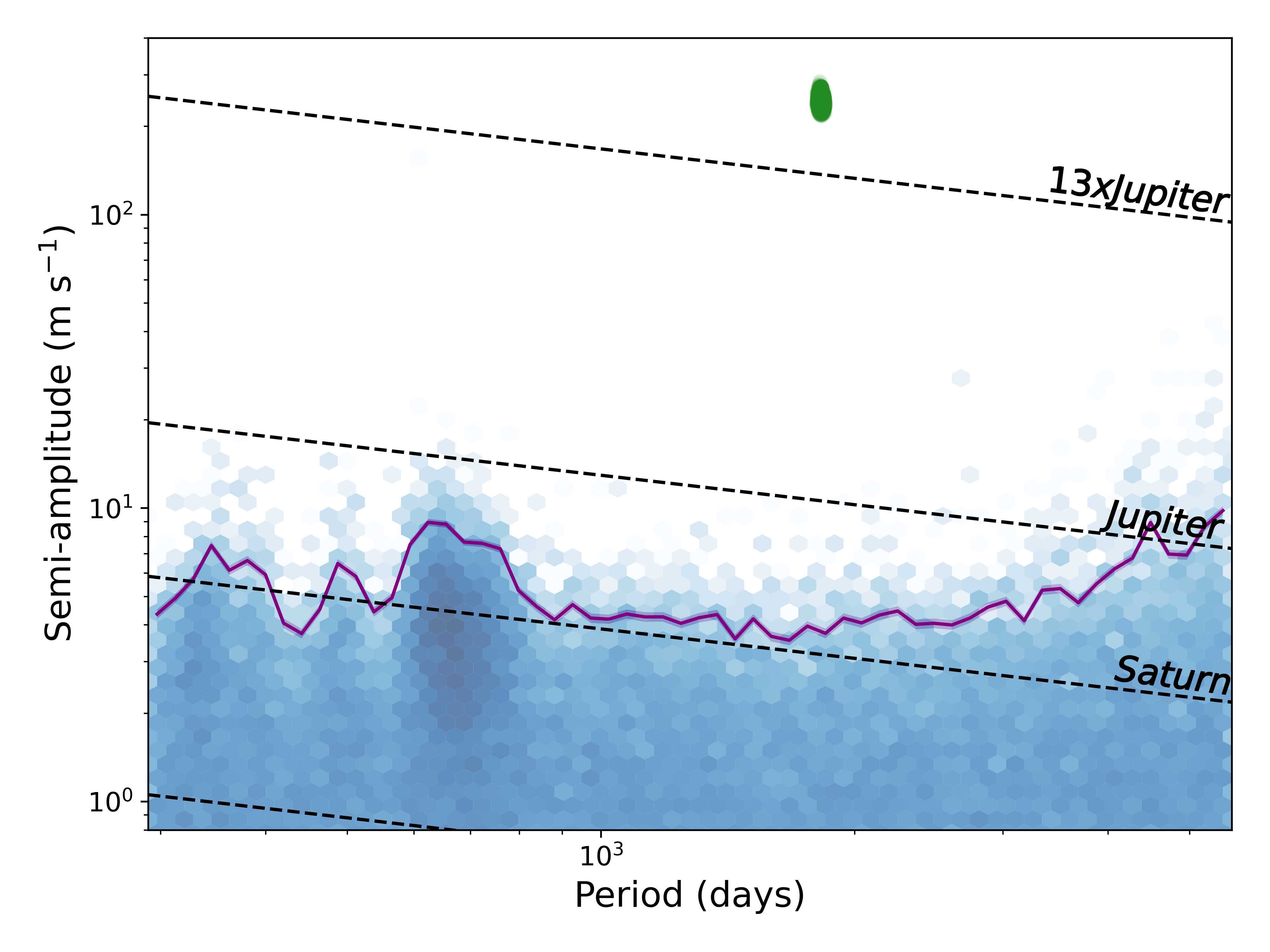}
    \caption{Detection limit plot for the BEBOP-4 system. In blue, the posterior density of the {\tt kima} run forced to fit a signal, with the brown-dwarf companion removed. The posterior distribution for the companion is shown in green. Dashed lines show expected signals of solar-system mass planets. The 99\% confidence detection limit as well as its uncertainty are shown in purple.}
    \label{fig:detlim}
\end{figure}

We calculate detection limits for any further signals in the radial velocity data in the same way as computed in  \citet{standing_bebop_2022,standing_radial-velocity_2023,baycroft25b}. The signal of the circumbinary companion is removed from the data, using the maximum-likelihood posterior sample, and the {\tt kima} analysis is re-run with the number of planets fixed to 1. The resulting posterior distribution corresponds to all the Keplerian signals that are consistent with the noise in the data, and a 99\% upper limit in semi-amplitude is calculated. The detection limit is shown in Fig. \ref{fig:detlim} along with the posterior distribution for the circumbinary companion for comparison. The uncertainty in the detection limit is $ E = \frac{1.97}{\sqrt{N}}$ where $E$ is the fractional uncertainty in the detection limit for a given period bin, and $N$ is the number of posterior samples in that bin. This relation is based on numerical experiments with {\tt kima} \citep{standing_radial-velocity_2023} and is similar to the relation used in \citet{standing_radial-velocity_2023} except they used a different scaling constant. From this exercise we confidently exclude the presence of any other planetary companion with a mass greater than Jupiter's.

Based on the current parameters, we calculate the detectability of BEBOP-4b with {\it Gaia}. Using the Gaia Observation Forecast tool\footnote{\href{https://gaia.esac.esa.int/gost/}{https://gaia.esac.esa.int/gost/}} we find that there will be up to 82 epochs of astrometric data published in {\it Gaia} DR4. Assuming a typical epoch precision of 40 \(\rm \mu as\) \citep{holl_gaia_2023}, we calculate the \textit{Gaia} SNR as in \citet{Sahlmann2015} in which a conservative \textit{Gaia} SNR $=20$ is recommended as a detection threshold. The BEBOP-4(AB) binary will have a \textit{Gaia} \({\rm SNR}\approx 100\), and BEBOP-4(AB)b a \textit{Gaia} \({\rm SNR}\approx 40\)\footnote{It is not expected that all of the epochs predicted will be present in the \textit{Gaia} data; for the SNR of the companion to get below 20 would require that only 24 of the 82 measurements are available.}. We therefore expect that \textit{Gaia} DR4 will be sensitive to the binary's orbit and the brown dwarf's orbit, allowing for measurements of the true mass and the mutual inclination. Amongst all confirmed circumbinary planets discovered with the transit and/or radial velocity methods, BEBOP-4\,b is currently the only one which is detectable with {\it Gaia}.

\section{Discussion}\label{sec:dis}

\subsection{Comparisons with other planets}

Compared to single star systems, BEBOP-4\,b does not particularly stand out. The exoplanet encyclopaedia \citep[\href{www.exoplanet.eu}{exoplanet.eu};][]{Schneider2011} lists 168 objects with mass $> 10\,\rm M_{jup}$ discovered either with radial velocities or transits, 24 of which orbit stars with masses $> 1.4\,\rm M_\odot$. 

Compared to known circumbinary planets identified with transits or radial velocities, BEBOP-4\,b is unique. It is the most massive by a factor $>3$.  In addition, BEBOP-4 both contains the longest binary period, and the largest planetary period in the known population. Thanks to its large semi-amplitude, it also has the most precise eccentricity, and is the most eccentric so far discovered. 

In terms of both mass and orbital parameters, BEBOP-4\,b is compatible with some of the proposed solutions for post common envelope binaries, where eclipse timing variations appear to imply a planetary companion. Some examples are: HW Vir which has claimed companions of comparable mass but much longer orbital periods than BEBOP-4b \citep{Beuermann_2012,baycroft_new_2023}; V1828 Aql \citep{Almeida_2013,Wolf_2021} and QS Vir \citep{Qian_2010,Basturk_2023} which have claimed companions of similar orbital periods but slightly lower masses\footnote{We note here that many claims for circumbinary exoplanets orbiting post common envelope binaries rely on eclipsing timing variations that may have other origins than being produced by a planetary body \citep[e.g.][]{Zorotovic2013,Hardy2015,baycroft_new_2023}.}. As such, BEBOP-4 might be seen as a precursor to the companions claimed orbiting post-common envelope binaries. The evolution of a population of circumbinary systems from main-sequence to circum-double-white-dwarf was performed by \citet{Columba_2023} who find that 20-30\% of  planets survive the whole evolution. Coincidentally, BEBOP-4A is itself  evolving off the main-sequence (near its terminal age main sequence), and about to ignite H-shell burning, on its way to become a sub-giant.

{\it Gaia}'s contribution in detecting new circumbinary exoplanets will mainly be if massive circumbinary planets $> 5\,\rm M_{jup}$ exist \citep[orbiting main sequence binaries, but also in post-common envelope systems;][]{Sahlmann2015}. BEBOP-4\,b reveals that some detections are to be expected.

Protoplanetary disc masses have been observed to scale with central mass \citep[e.g.][]{Pascucci2016,Andrews2018} and the combined stellar mass at the centre of the system is $M_{\rm A} + M_{\rm B} = 1.97\, \rm M_\odot$, uncommonly large for most planets orbiting single stars, but a fairly regular occurrence for circumbinary planets \citep[e.g. TIC 172900988;][]{kostov_tic_2021,sairam_new_2024}.  As such the BEBOP-4 system is a good example of why studying circumbinary planets will support a better understanding on how planet formation scales with central mass. Whilst detecting a planet orbiting a single $2\,\rm M_\odot$ star is difficult (lack of absorption lines, and fast rotation), doing so by sharing the mass between two stars makes for, in this particular case, an easier detection.



\subsection{Possible formation pathways}
As mentioned in Sect.~\ref{sec:intro}, the occurrence frequency of brown dwarf companions to main-sequence stars has a distinct minimum for masses between $\sim 25-45\,\rm M_{jup}$, with the occurrence rate increasing both towards the hydrogen burning limit at $\sim 80\,\rm M_{jup}$ and towards the deuterium burning limit at $\sim 13\,\rm M_{jup}$ \citep[e.g.][]{Grether2006, Sahlmann2014, Dalal2021}. This bimodality suggests two formation channels are at play, with higher-mass brown dwarfs being the low-mass tail of the stellar population that forms through fragmentation of molecular cloud cores, and lower-mass objects being the high-mass end of planet formation in protoplanetary discs. Based on its minimum mass, and the constraint on its true mass from the binary's apsidal precession, BEBOP-4b appears to be an example of the latter population that formed in a circumbinary disc.

The discussion below about possible formation scenarios assumes that during the epoch of BEBOP-4b's formation, a circumbinary disc is present around the BEBOP-4 binary system, and the binary has component masses and orbital elements that are similar to those observed today. Consideration of the earlier evolution goes beyond the scope of this discussion. A further point to note is that the central binary contains a total mass $\sim 2$~M$_{\odot}$ and disc masses are known to scale at least linearly with the central stellar mass \citep[e.g.][]{Pascucci2016}. Hence, we assume the circumbinary disc contains sufficient mass to form a planet as massive as BEBOP-4b.

Formation in a circumbinary disc could have occurred through fragmentation of the disc \citep[e.g.][]{StamatellosWhitworth2009} or core accretion \citep{Pollack1996, coleman_global_2023}. 
In the disc fragmentation scenario, the cooling time in the disc needs to be short and on the order of the local orbital time for a gravitationally unstable disc to form one or more bound clumps. Hence, fragmentation is expected in the outer regions of a protoplanetary disc beyond $\sim 50$~au \citep[e.g.][]{Zhu2012}. Given that BEBOP-4b has a semi-major axis of $a_{\rm p} = 3.63$~au, formation through disc fragmentation requires it to have migrated inwards to its currently observed location. This is generally expected, since simulations of clumps formed by disc fragmentation find that disc-driven inwards migration can occur rapidly \citep{Boley2010,Baruteau2011,Zhu2012}. In their study, \citet{Zhu2012} showed that clumps that grow massive enough to open gaps in the disc dramatically slow their migration. Hence, a plausible scenario for the origin of BEBOP-4b is formation at large radius in a massive circumbinary disc followed by inwards migration to its current location. However, caution should be taken. \citet{Zhu2012} considered a coarsely sampled set of initial conditions, and the objects that opened gaps and survived migration were more massive than BEBOP-4b. Furthermore, a recent study aimed at understanding the origins of the Delorme 1 circumbinary planet showed that inwards migration of the planet is often reversed through interaction with a gravitationally unstable disc \citep{TeasdaleStamatellos2024}. Hence, a tailored study will be required to determine whether the above scenario for BEBOP-4b's formation is actually viable.

Comparing observations of transiting planets with masses in the range $20$~M$_{\oplus} < M_{\rm p} < 20$ M$_{\rm jup}$ around single stars with planetary interior models demonstrates that gas giant planets are enriched with heavy elements, and the degree of enrichment depends on the mass of the planet, consistent with expectations from the core accretion scenario for planet formation \citep{MillerFortney2011,ThorngrenFortney2016}. Hence, it seems highly probable that at least some of the observed massive super-Jovian planets formed via core accretion. If formation occurs sufficiently far from the central binary, where its gravitational perturbations are small enough that pebble and planetesimal accretion are not inhibited \citep{paardekooper_how_2012, pierens_vertical_2021}, then it is plausible that circumbinary super-Jovian planets such as BEBOP-4b can also form through this mechanism, with their final orbital locations being determined by post-formation disc-driven migration \citep{pierens_formation_2008} and the timing of disc dispersal. The formation of such a massive planet faces significant obstacles, which may explain why these bodies are rare compared to lower-mass planets. For example, gas giant planets open gaps in their protoplanetary discs and for planets that exceed Jupiter's mass, efficient tidal truncation of the disc and deep gap formation can reduce the gas accretion rate onto the planet, making it difficult to accumulate a sufficiently massive gas envelope within the disc lifetime \citep[e.g.][]{Bryden1999,Nelson2023}. However, as a planet grows beyond $\sim 3$~M$_{\rm jup}$, both the disc and the planet's orbit can start to become eccentric \citep{Papaloizou2001,KleyDirksen2006,Ragusa2018}, reducing the effectiveness of gap opening in quenching gas accretion onto the planet and providing the possibility of massive gas envelopes accreting on Myr timescales. This disc-driven eccentricity growth may also help to explain the observed eccentricity of BEBOP-4b, although the inferred value of $e_{\rm b} =  0.43$ is larger than the values of $e \sim 0.3$ that have been obtained in hydrodynamical simulations to date. However, larger values of $e_{\rm b}$ might be achieved by varying the conditions in the disc such as the temperature and viscosity. The driving of eccentricity through disc interaction is attractive when considering the probability that BEBOP-4b is coplanar with the central binary. The most likely alternative scenario for exciting BEBOP-4b's eccentricity is planet-planet scattering, and this would be expected to increase both the eccentricity and the inclination relative to the binary orbit plane, reducing the likelihood for a transit to be detected. Finally, we note that unlike the transit-discovered circumbinary planets, BEBOP-4b is not orbiting especially close to the instability zone near the binary (see Fig.~\ref{fig:orbit}), and hence does not obviously fit the picture considered by \citet{pierens_formation_2008}
 and \citet{penzlin_parking_2021} of having migrated from further out and then being parked at the edge of the cavity formed by the binary. However, the large mass of BEBOP-4b, combined with the likelihood it experienced significant disc-driven eccentricity excitation, likely led to significant sculpting of the disc, resulting in BEBOP-4b and the central binary orbiting within a common cavity. Future simulations will be required to determine what the implications of this would have been for the evolution of BEBOP-4b.



\subsection{Orbital stability in the system}

The proximity of a circumbinary planet's orbit to the stability limit can be formulated in a variety of ways, which can lead to different interpretations. We use the parameter \(a_{\rm sc} = a_{\rm b}/r_{\rm stab}\) \citep{baycroft25b}, where \(a_{\rm b}\) is the semi-major axis of the planet's orbit and \(r_{\rm stab}\) is the radius of the instability zone/the critical semi-major axis. Using the formula for the critical semi-major axis from \citet{holman_long-term_1999} which takes into account \(a_{\rm bin}\), \(e_{\rm bin}\), and \(q_{\rm bin}\) gives a value of \(a_{\rm sc} \approx 2.7\). Instead using the formula from \citet{Georgakarakos2024}, which includes more parameters (notably the eccentricity of the planet's orbit) results in \(a_{\rm sc} \approx 1.3 - 1.7\) \citep[the two values being the inner and outer stability criteria defined in][]{Georgakarakos2024}. Instead of \(a_{\rm sc}\) using the \citet{holman_long-term_1999} formula, we can incorporate the eccentricity of the brown dwarf's orbit by calculating \(r_{\rm peri,sc} = \frac{r_{\rm peri,b}}{r_{\rm stab}}\), where \(r_{\rm peri,b}\) is the pericentre distance of a circumbinary planet's orbit. This yields a value of \(r_{\rm peri,sc} \approx 1.5\). BEBOP-4b can therefore be considered close or far from the stability limit depending on how its eccentricity is dealt with.

To get a more definitive view about the stability of the current system, we performed a global dynamical analysis in the same way as achieved for other circumbinary planetary systems \citep[eg.][]{correia_coralie_2005, standing_radial-velocity_2023}.
We used the symplectic integrator SABAC4 \citep{laskar_high_2001}, with a step size of $0.005$~yr and general relativity corrections.
Each initial condition is integrated for $100$~kyr, and a stability indicator, $\Delta = |1-n'/n|$, is computed, where $n$ and $n'$ are the main frequency of the mean longitude over two consecutive time intervals of $50$~kyr, calculated via frequency analysis \citep{laskar_chaotic_1990, laskar_frequency_1993}. 
The results are reported in color, where yellow represents strongly chaotic trajectories with $\Delta > 10^{-2}$, while extremely stable systems for which $\Delta < 10^{-8}$ are shown in purple/black. 
Orange indicates the transition between the two, with $\Delta \sim 10^{-4}$.

\begin{figure}
    \centering
    \includegraphics[width=\linewidth]{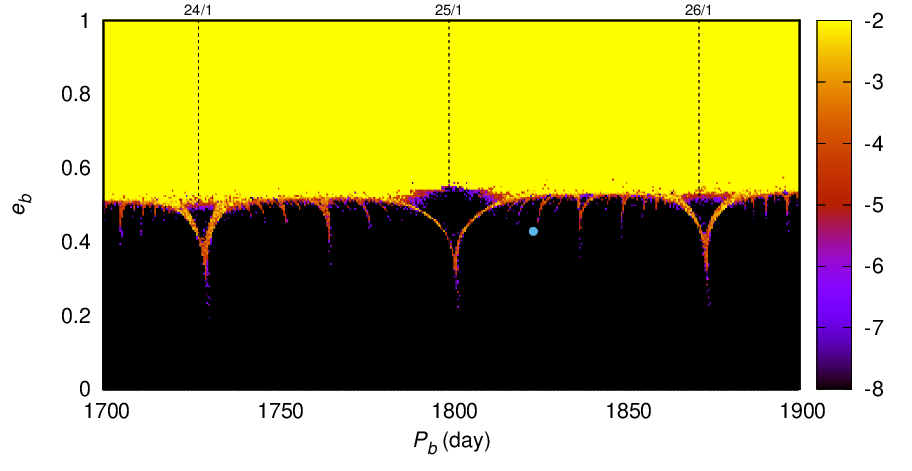}
    \caption{Stability analysis of the brown-dwarf in the BEBOP-4 system assuming coplanar orbits. For fixed initial conditions (Tabs.~\ref{tab:parameters}), the parameter space of the system is explored by varying the orbital period $P_\mathrm{inner}$ and the eccentricity $e_\mathrm{inner}$ of the brown-dwarf. The step size is $0.5$~day in orbital period and $0.005$ in eccentricity. For each initial condition, the system is integrated over $10^5$~yr and a stability indicator is calculated which involved a frequency analysis of the mean longitude of the ibrown-dwarf. Chaotic diffusion is indicated when the main frequency of the mean longitude varies in time. Yellow points correspond to highly unstable orbits, while purple points correspond to orbits which are likely to be stable on a billion-years timescales. The dashed black lines indicate mean-motion resonances with the binary stars. The blue dot gives the position of the best fit solution (Tabs.~\ref{tab:parameters}).}
    \label{fig:stability}
\end{figure}

\begin{figure}
    \centering
    \includegraphics[width=\linewidth]{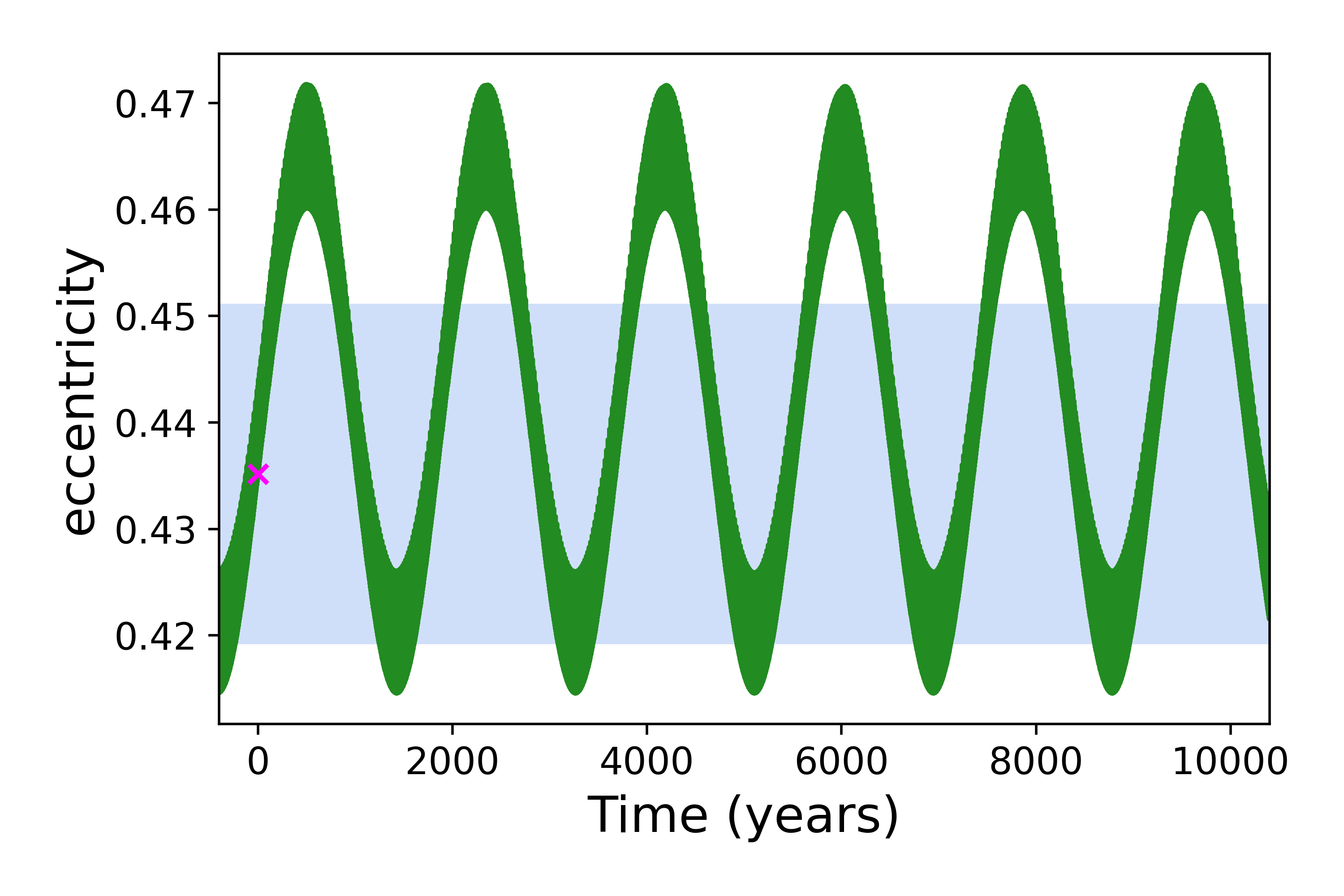}
    \caption{Evolution of the eccentricity of BEBOP-4b. The system is initiated with the maximum-likelihood parameters from the {\tt kima} analysis, the eccentricity of which (\(e = 0.435\)) is marked. The uncertainty in the measured eccentricity is shown (centered on the maximum-likelihood value) as a shaded region.}
    \label{fig:eevolution}
\end{figure}

In Figure~\ref{fig:stability}, we vary the orbital period and the eccentricity of the brown-dwarf in the vicinity of the best fit (Table~\ref{tab:parameters}).
We observe that that the system is stable up to eccentricities around 0.5, which is very close to the observed eccentricity ($e=0.435$). 
Moreover, there are some high order mean motion resonances with the binary nearby, namely the 25/1, that also introduce some chaotic structures for lower eccentricities.
We conclude that the the best fit system (given by a blue dot) is stable, but indeed near its stability limit, which raises interesting questions on how this system has formed.
In Figure~\ref{fig:eevolution}, we show how the eccentricity of the brown dwarf's orbit evolves over time. The amplitude of the variations in eccentricity exceed the statistical uncertainty on the measured value of the mean eccentricity from the {\tt kima} analysis. 


A recent comparison between the transit-detected circumbinary planets, and the radial-velocity detected circumbinary planets shows tension in their distribution relative to the stability limit \citep{holman_long-term_1999,Pilat2003,Georgakarakos2024}. Whilst the innermost, transiting planets are almost always detected near the stability limit \citep{martin_planets_2014,li_uncovering_2016}, radial-velocity system do not seem to show such preference \citep{Baycroft24}. One hypothesis to explain this discrepancy is for radial-velocity detected systems to host an as-of-yet undetected, low mass (but large radius) exoplanet near the stability limit. In this context, we study here whether there are dynamically stable locations between the binary and the brown dwarf for an additional planet to subsist, which might inform further data collections.

In order to check if additional planets can be hidden in the inner system, we performed again a global dynamical analysis.
We assumed the best fit solution for the already known bodies in the system (Table~\ref{tab:parameters}), and scanned all orbital periods and eccentricities of a tentative inner coplanar planet between the binary stars and  the brown-dwarf with a radial-velocity signature of 5~\({\rm m\,s^{-1}}\).
We observed that all orbits are unstable, and thus conclude that no other planet can subsist in the inner system.

\section{Concluding remarks}
Thanks to 84 spectra obtained with the SOPHIE high-resolution spectrograph, we discovered BEBOP-4b, a planetary-mass brown dwarf with $m_{\rm b}\sin{i_{\rm b}} = 20.9\pm1.3 \,\rm M_{jup}$, the most massive amongst circumbinary planets confirmed via the transit or radial velocity methods. Based on the binary's apsidal precession we also expect the true mass remains $<26.3\,\rm M_{jup}$. The circumbinary brown dwarf occupies a $1823.5^{+5.1}_{-4.9}\,\rm d$ orbit, with a large eccentricity $e = 0.428\pm0.016$, the highest yet measured amongst circumbinary planets. The binary has masses $M_A = 1.51\,\rm M_\odot$, and $M_B = 0.46\,\rm M_\odot$, an orbital period $P_{\rm bin} = 72\,\rm d$ and $e_{\rm bin} = 0.27$. BEBOP-4\,AB is the longest period binary known to host a circumbinary planet detected with transit or radial-velocities. At this stage, BEBOP-4\,(AB)\,b is unremarkable when compared to known exoplanets orbiting single stars, but unusual within the circumbinary context. Because planet formation is expected to be affected by binarity, it is surprising that such a high mass object could assemble. As surveys for circumbinary exoplanets expand their monitoring, and thanks to {\it Gaia}, it will soon be possible to know whether such high-mass companions are a regular occurrence.

\section*{Acknowledgements}

These scientific results would not have been possible without a generous allocation of telescope time by France Programme National de Planétology, and without the dedication of staff working at the Observatoire de Haute-Provence, particularly during the COVID-19 pandemic. Four measurement were obtained during that period, which are essential for our conclusions. We  would also like to thank the help of other observers part of the timeshare agreement between groups using the SOPHIE instrument. 
This research leading to these results is supported by a grant from the European Research Council (ERC) under the European Union's Horizon 2020 research and innovation programme (grant agreement n$^\circ$ 803193/BEBOP) and by .
ACMC acknowledges support from FCT - Funda\c{c}\~ao para a Ci\^encia e a Tecnologia, I.P., Portugal, through the CFisUC projects UIDB/04564/2020 and UIDP/04564/2020, with DOI identifiers 10.54499/UIDB/04564/2020 and 10.54499/UIDP/04564/2020, respectively. Computations described in this paper were performed using the University of Birmingham's BlueBEAR HPC service.
The stability maps were performed at the Oblivion Supercomputer at the University of \'Evora (\href{https://oblivion.hpc.uevora.pt}{https://oblivion.hpc.uevora.pt}).
\section*{Data Availability}

The {\it TESS} data is available to download via the MAST portal \href{https://mast.stsci.edu}{https://mast.stsci.edu}. Spectroscopic data are available on the \href{http://atlas.obs-hp.fr/sophie/}{SOPHIE archive}, and were obtained under Prog.ID 18B.DISC.TRIA and 19A.PNP.SANT, and under target name J1250+26. Reduced and drift-corrected radial velocities are found in Table \ref{tab:rv_data}, a full version of which can be found online.



\bibliographystyle{mnras}
\bibliography{bebop4} 




\appendix

\section{Additional Figures and Tables}

\begin{figure}
    \centering
    \includegraphics[width=1.0\linewidth]{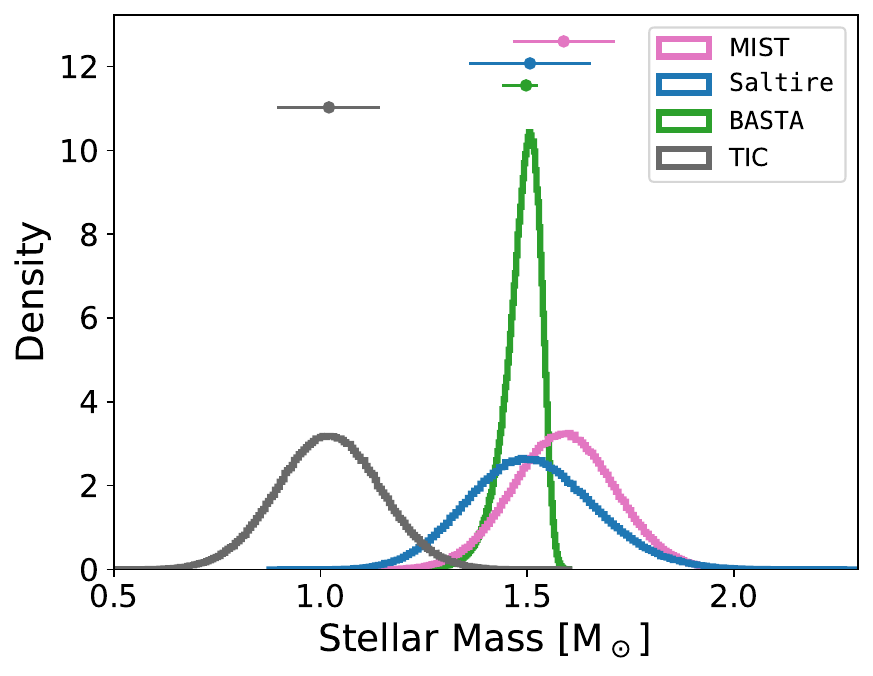}
    \caption{A comparison of the mass estimates for BEBOP-4\,A. The results from Section~\ref{sec:M1} and the HRCCS analysis with \texttt{Saltire} are consistent with inference from \texttt{BASTA} from the observed [Fe/H], $T_{\rm eff}$, and G magnitude using the BaSTI isochrones \citep{Hidalgo2018} with micro-physics following the recommended science case for this mass range, and with  the MIST isochrone value reported by \citet{freckelton_bebop_2024} for BEBOP primary stars. The \textit{TESS} input catalog (TIC) mass is almost fully inconsistent with other estimates (with a discrepancy in the median of $\sim0.5\,\rm{M}_{\odot}$), highlighting the need to derive accurate stellar parameters. Mock distributions are randomly generated from values and their $1\sigma$ uncertainties to visualize the degree of consistency between estimates.}
    \label{fig:mass-comparison}
\end{figure}

\begin{figure*}
    \centering
    \includegraphics[width=\linewidth]{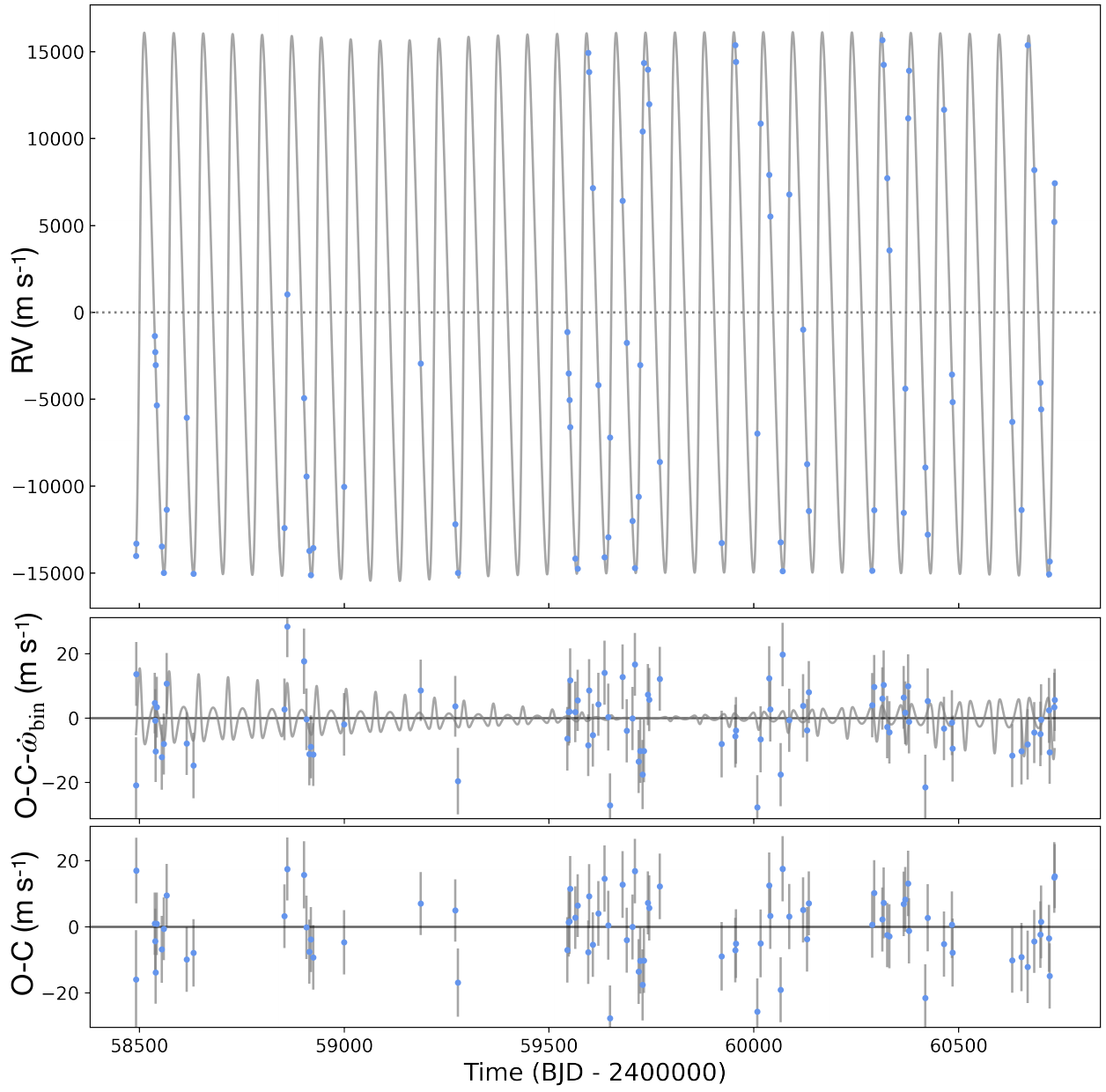}
    \caption{Top panel: Radial velocity observations (in blue) as a function of time, with the highest likelihood model represented (grey line). Middle panel: residuals after removing the highest likelihood model except for $\dot{\omega}_{\rm bin}$ fixed to 0. The grey line the contribution of a non zero $\dot{\omega}_{\rm bin}$ to the fit. Bottom plot: residuals after removing the highest likelihood model (including $\dot{\omega}_{\rm bin}$).}
    \label{fig:binary_wdot}
\end{figure*}

\begin{figure}
    \centering
    \includegraphics[width=\linewidth]{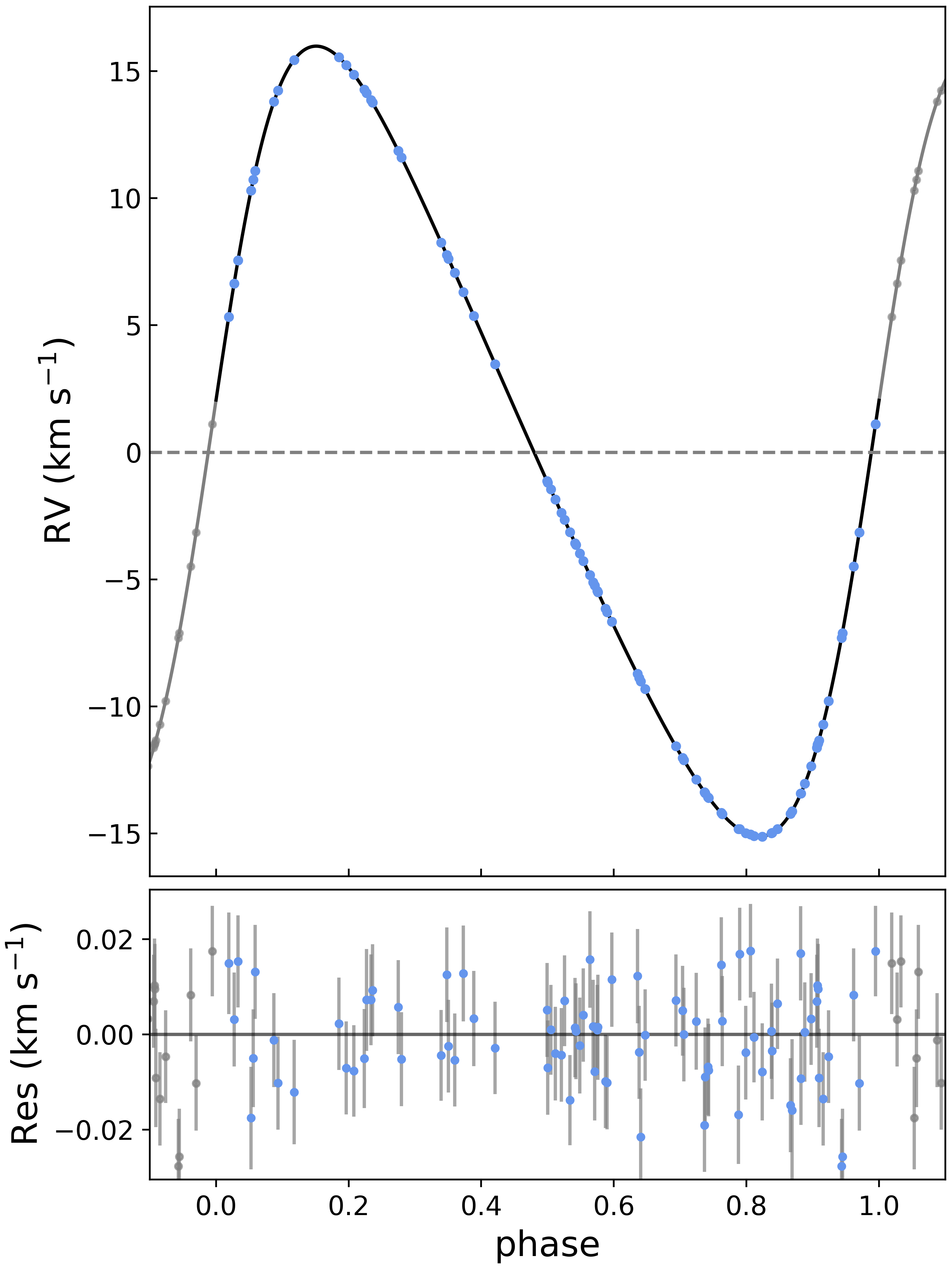}
    \caption{Radial velocity Keplerian for the binary star (secondary's orbit around the primary). Residuals (including companion removed) are shown below.}
    \label{fig:binary_rv}
\end{figure}

\begin{figure}
    \centering
    \includegraphics[width=\linewidth]{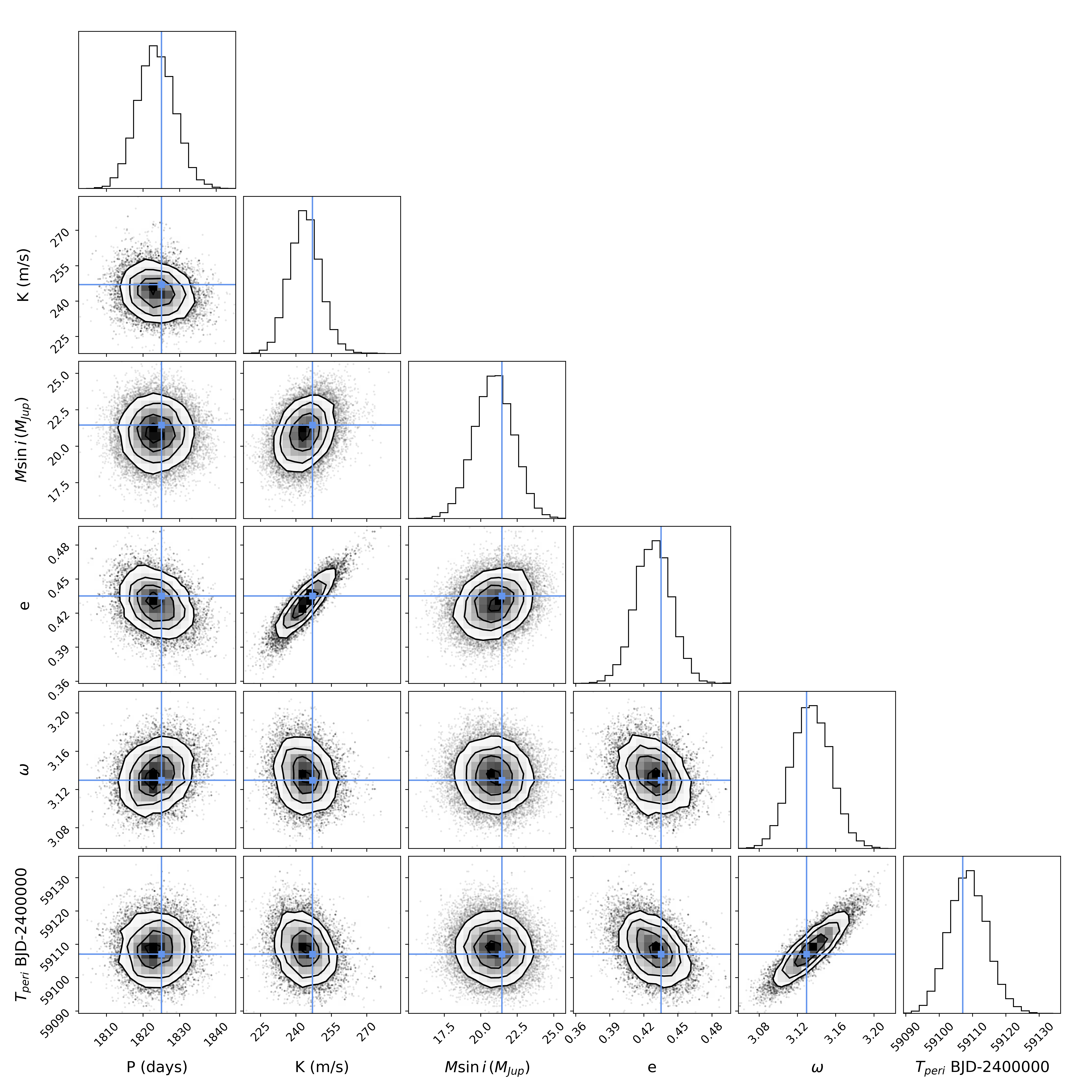}
    \caption{Distributions and correlations of the brown dwarf's parameters. The lines show the values of the maximum-likelihood solution which is used to generate the orbital and radial velocity phase plots as well as to instantiate the N-body simulations.}
    \label{fig:planet_corner}
\end{figure}

\begin{table}
    \centering
     \caption{Radial velocity observations taken with SOPHIE between 2019-01-07 and 2025-02-27. Full table can be found in online version of paper.}
    \begin{tabular}{ccccc}
    \hline
    Time & Vrad & \(\sigma{\rm -Vrad}\) & FWHM & Bisector Span \\
    \([{\rm BJD} - 2400000]\) & [\(\rm km\,s^{-1}\)] & [\(\rm km\,s^{-1}\)] &[\(\rm km\,s^{-1}\)] & [\(\rm km\,s^{-1}\)] \\
    \hline
    58491.71259 & -8.73180 & 0.01170 & 10.62970 & -0.05000 \\
    58492.64226 & -8.03090 & 0.00350 & 10.44850 & 0.00190 \\
    58537.49195 & 3.92270 & 0.00180 & 10.49780 & 0.00370 \\
    58538.63542 & 2.99850 & 0.00330 & 10.47860 & -0.01230 \\
    58539.57288 & 2.24080 & 0.00200 & 10.46250 & 0.00260 \\
    58542.53512 & -0.07170 & 0.00190 & 10.44340 & -0.00160 \\
    58554.56388 & -8.18770 & 0.00410 & 10.45980 & 0.00050 \\
    58559.53582 & -9.71920 & 0.00210 & 10.48910 & 0.00480 \\
     58566.48890 & -6.06990 & 0.00180 & 10.49830 & 0.00790 \\
     58615.39100 & -0.78710 & 0.00320 & 10.44890 & -0.00870 \\
     58632.41799 & -9.76290 & 0.00430 & 10.44930 & 0.01950 \\
     58853.61602 & -7.12970 & 0.00260 & 10.45000 & -0.00850 \\
     58860.59606 & 6.31790 & 0.00220 & 10.50650 & 0.01910 \\
     58901.58528 & 0.34410 & 0.00410 & 10.47490 & 0.01670 \\
     58907.55645 & -4.15700 & 0.00250 & 10.46640 & 0.00660 \\
     58914.46217 & -8.44100 & 0.00290 & 10.48530 & 0.00810 \\
     58918.47879 & -9.83800 & 0.00330 & 10.48370 & 0.01410 \\
     58924.49174 & -8.27870 & 0.00310 & 10.46700 & 0.00520 \\
     58999.45405 & -4.75600 & 0.00300 & 10.53470 & 0.00710 \\
     59186.70116 & 2.33260 & 0.00230 & 10.44310 & -0.00990 \\
     59271.47422 & -6.90930 & 0.00190 & 10.53520 & 0.00710 \\
     59277.53426 & -9.70900 & 0.00460 & 10.50070 & 0.03490 \\
     59544.70327 & 4.14610 & 0.00350 & 10.47750 & 0.00810 \\
     59547.69428 & 1.76270 & 0.00480 & 10.40200 & 0.01320 \\
     59549.65688 & 0.22340 & 0.00330 & 10.47790 & 0.00510 \\
     59551.68819 & -1.32700 & 0.00350 & 10.45920 & -0.00450 \\
     59563.65668 & -8.88740 & 0.00190 & 10.47810 & 0.00240 \\
     59569.65054 & -9.47360 & 0.00220 & 10.48150 & -0.00790 \\
     59595.64463 & 20.21810 & 0.00260 & 10.45190 & 0.00040 \\
     59597.67188 & 19.11170 & 0.00270 & 10.43870 & 0.01470 \\
     59606.58607 & 12.42750 & 0.00310 & 10.40990 & 0.02790 \\
     59620.55505 & 1.08840 & 0.00320 & 10.45620 & -0.00040 \\
     59635.52895 & -8.81210 & 0.00370 & 10.46940 & 0.01040 \\
     59644.60100 & -7.66070 & 0.00480 & 10.43150 & -0.02130 \\
     59648.58000 & -1.92710 & 0.00360 & 10.44000 & -0.00480 \\
     59679.50840 & 11.68760 & 0.00390 & 10.43210 & 0.01520 \\
     59689.47297 & 3.52490 & 0.00330 & 10.42980 & 0.00970 \\
     59703.42891 & -6.72190 & 0.00330 & 10.39760 & 0.00620 \\
     59709.49403 & -9.43830 & 0.00300 & 10.45750 & -0.01370 \\
     59718.53782 & -5.32550 & 0.00320 & 10.45230 & 0.02460 \\
     59722.46460 & 2.24050 & 0.00370 & 10.40180 & 0.00450 \\
     59728.42578 & 15.68930 & 0.00550 & 10.45860 & -0.01920 \\
     59731.36069 & 19.62520 & 0.00310 & 10.39960 & -0.01480 \\
     59741.44020 & 19.25100 & 0.00230 & 10.38600 & 0.01280 \\
     59744.40568 & 17.25560 & 0.00340 & 10.45150 & 0.00240 \\
     59770.35912 & -3.31800 & 0.00350 & 10.43350 & -0.00220 \\
     59921.65526 & -7.99220 & 0.00470 & 10.47920 & 0.01940 \\
     59954.66702 & 20.65160 & 0.00310 & 10.45480 & 0.00200 \\
     59956.62189 & 19.68750 & 0.00470 & 10.48190 & -0.05370 \\
     60008.51670 & -1.69720 & 0.00390 & 10.46720 & 0.02930 \\
     60016.54193 & 16.14430 & 0.00450 & 10.42250 & -0.01300 \\
     60037.56070 & 13.18390 & 0.00360 & 10.43360 & 0.00070 \\
     60040.48115 & 10.78860 & 0.00380 & 10.43670 & 0.01180 \\
     60065.49599 & -7.95010 & 0.00330 & 10.51490 & 0.00050 \\
     60070.48927 & -9.61340 & 0.00330 & 10.47600 & -0.02240 \\
     60086.42507 & 12.06260 & 0.00340 & 10.40860 & -0.01010 \\
     60120.39316 & 4.28410 & 0.00340 & 10.46250 & 0.00440 \\
     60130.39260 & -3.45480 & 0.00300 & 10.45290 & -0.01330 \\
     60134.37139 & -6.14830 & 0.00280 & 10.46450 & -0.02530 \\
     60288.68283 & -9.58510 & 0.00380 & 10.46000 & -0.00400 \\
     60293.69433 & -6.10210 & 0.00350 & 10.48120 & -0.00320 \\
     60313.69696 & 20.94420 & 0.00290 & 10.38330 & 0.01330 \\
     60316.71702 & 19.53310 & 0.00530 & 10.36360 & 0.00710 \\
     60325.60521 & 13.00350 & 0.00300 & 10.43340 & -0.01030 \\
     60330.66027 & 8.85680 & 0.00290 & 10.42120 & 0.00880 \\
     60365.56386 & -6.24620 & 0.00310 & 10.48670 & 0.01440 \\
     60369.58676 & 0.88660 & 0.00310 & 10.46640 & 0.04220 \\
     60376.58372 & 16.45240 & 0.00340 & 10.39050 & -0.00590 \\
     60378.62953 & 19.18100 & 0.00340 & 10.38890 & 0.01220 \\
     60418.44175 & -3.64580 & 0.00420 & 10.44800 & -0.03590 \\
     60424.45937 & -7.51300 & 0.00420 & 10.40280 & 0.03970 \\
     60464.44600 & 16.94550 & 0.00340 & 10.43320 & 0.00200 \\
     60483.37189 & 1.69910 & 0.00400 & 10.39550 & -0.04270 \\
     60485.39048 & 0.10770 & 0.00450 & 10.40930 & 0.00940 \\
     60630.68770 & -1.02540 & 0.00320 & 10.44540 & -0.01620 \\
     60653.67890 & -6.09260 & 0.00460 & 10.45900 & -0.01090 \\
     60668.67833 & 20.66010 & 0.00590 & 10.40810 & -0.03850 \\
     60684.65787 & 13.46230 & 0.00220 & 10.42710 & 0.01000 \\
     60699.68725 & 1.22490 & 0.00390 & 10.42510 & -0.01110 \\
     60701.65134 & -0.30700 & 0.00600 & 10.46830 & 0.00620 \\
     60720.56465 & -9.79630 & 0.00410 & 10.46480 & -0.01800 \\
     60722.55497 & -9.05170 & 0.00340 & 10.47250 & -0.00110 \\
     60733.55300 & 10.48150 & 0.00530 & 10.40760 & -0.03590 \\
     60734.55731 & 12.70360 & 0.00270 & 10.41540 & -0.00870 \\
    \hline
    \end{tabular}
    \label{tab:rv_data}
\end{table}


\begin{landscape}

\begin{table}
\centering
\caption{The BEBOP catalogue of circumbinary planets with radial velocity measurements. \(1\sigma\) uncertainties on parameters are provided at the level of the last two significant digits. \(^x\) Transit detection from \citet{doyle_kepler-16_2011} .\(^y\) Transit detection and parameters from \citet{kostov_toi-1338_2020} .\(^z\) Transit detection from \citet{kostov_tic_2021}.}
\begin{tabular}{l|c|c|c|c|c|c|c|c|c|r}
     Catalogue entry & Other names & \(P_{\rm bin}\) & \(M_{\rm pri}\) & \(M_{\rm sec}\) & \(e_{\rm bin}\) & \(P_{\rm pl}\) & \(M_{\rm pl}\) & \(e_{\rm pl}\)  & Source of RVs & TIC ID \\
      &  & [days] & [\(M_{\rm \odot}\)] & [\(M_{\rm \odot}\)] &  & [days] & [\(M_{\rm jup}\)] &  &  &   \\
    \hline
     BEBOP-0 b & Kepler-16 b\(^x\) & \(41.077772(51)\) & \(0.654(17)\) & \(0.1964(31)\) & \(0.15994(10)\) & \(226.0(1.7)\) & \(0.313(39)\) & \(\leq0.21\) & \citet{triaud_bebop_2022} & 299096355\\
     BEBOP-1 b & TOI-1338 b\(^y\)  & \(14.6085579(57)\) & \(1.098(17)\) & \(0.307(3)\) & \(0.155522(29)\) & \(95.174(35)^y\) & \(<0.0685(29)\) & \(0.0880(43)^y\) & \citet{standing_radial-velocity_2023} & 260128333\\
     BEBOP-1 c & TOI-1338 c & \(14.6085579(57)\) & \(1.098(17)\) & \(0.307(3)\) & \(0.155522(29)\) & \(215.5(3.3)\) & \(0.205(37)\) & \(\leq0.16\) & \citet{standing_radial-velocity_2023} & 260128333 \\
     BEBOP-2 b & TIC 172900988 b\(^z\) & \(19.657878(34)\) & \(1.23681(39)\) & \(1.20207(33)\) & \(0.448234(90)\) & \(151.2(1.8)\) & \(1.90(25)\) & \(\leq0.11\) & \citet{sairam_new_2024} & 172900988 \\
     BEBOP-3 b & -- & \(13.2176657(27)\) & \(1.083(26)\) & \(0.3615(39)\) & \(0.063255(54)\) & \(547.0^{(+6.2)}_{(-7.6)}\) & \(0.558^{(+51)}_{(-48)}\) & \(0.247^{(+77)}_{(-89)}\) & \citet{baycroft25b} & 289949453 \\
     BEBOP-4 b & HD\,111065\,b& \(71.96858^{(+13)}_{(-15)}\) & \(1.507(98)\) & \(0.459(18)\) & \(0.26887(11)\) & \(1823.5(5.0)\) & \(21.56(83)\) & \(0.428(16)\) & This work & 356710041 \\
\end{tabular}
\label{tab:catalogue}
\end{table}
\end{landscape}


\bsp	
\label{lastpage}
\end{document}